\documentclass[openacc]{rstransa}%%%%where rstrans is the template name

\titlehead{Review}

\jname{rsta}
\Journal{Phil. Trans. R. Soc}

\begin{document}

%%%% Article title to be placed here
\title{Self-gravitating Superfluids: The Gross-Pitaevskii-Poisson Framework}

\author{%%%% Author details
Sanjay Shukla$^{1}$, Marc E. Brachet$^{2}$ and Rahul Pandit$^{3}$}

%%%%%%%%% Insert author address here
\address{$^{1}$Applied Physics and Science Education, Fluids and Flows, TU/e Eindhoven University of Technology, Het Eeuwsel 53, 5612 AZ Eindhoven, Netherlands\\
$^{2}$Laboratoire de Physique de l’\'Ecole Normale Supérieure, ENS, Universit\'e PSL,
CNRS, Sorbonne Universit\'e, Universit\'e de Paris, 24 Rue Lhomond, 75005 Paris, France\\
$^{3}$Centre for Condensed Matter Theory, Department of Physics, Indian Institute of Science,
Bangalore 560012, India}

%%%% Subject entries to be placed here %%%%
\subject{xxxxx, xxxxx, xxxx}

%%%% Keyword entries to be placed here %%%%

%%%% Insert corresponding author and its email address}
\corres{Rahul Pandit\\
\email{rahul@iisc.ac.in}}

%%%% Abstract text to be placed here %%%%%%%%%%%%
\begin{abstract}
We provide an overview of the Gross-Pitaevskii-Poisson equation (GPPE) that is used to model self-gravitating 
superfluid systems, which include gravitationally collapsed boson and axion stars and dark-matter haloes.
We outline how this framework can be used to develop minimal models for neutron stars and for pulsars and their glitches. We account not only for vortices in the neutron superfluid inside these stars, but also for the flux tubes in the proton-superconductor subsystem, using a coupled model with the neutron superfluid, proton superconductor, the Maxwell equations for the vector potential ${\bf A}$, and the Poisson equation for self-gravity.
\end{abstract}
%%%%%%%%%%%%%%%%%%%%%%%%%%%

%%%%%%%%%% Insert the texts which can accomdate on firstpage in the tag "fmtext" %%%%%

\begin{fmtext}
.....

%\section{Equations}
%%% Numbered equation
%\begin{align}\label{1.1}
%\begin{split}
%\frac{\partial u(t,x)}{\partial t} &= Au(t,x) \left(1-\frac{u(t,x)}{K}\right)-B\frac{u(t-\tau,x) w(t,x)}{1+Eu(t-\tau,x)},\\
%\frac{\partial w(t,x)}{\partial t} &=\delta \frac{\partial^2w(t,x)}{\partial x^2}-Cw(t,x)+D\frac{u(t-\tau,x)w(t,x)}{1+Eu(t-\tau,x)},
%\end{split}
%\end{align}

%\begin{align}\label{1.2}
%\begin{split}
%\frac{dU}{dt} &=\alpha U(t)(\gamma -U(t))-\frac{U(t-\tau)W(t)}{1+U(t-\tau)},\\
%\frac{dW}{dt} &=-W(t)+\beta\frac{U(t-\tau)W(t)}{1+U(t-\tau)}.
%\end{split}
%\end{align}

%%%% Unnumbered equation
%\begin{eqnarray}
%\frac{\partial(F_1,F_2)}{\partial(c,\omega)}_{(c_0,\omega_0)} = \left|
%\begin{array}{ll}
%\frac{\partial F_1}{\partial c} &\frac{\partial F_1}{\partial \omega} \\\noalign{\vskip3pt}
%\frac{\partial F_2}{\partial c}&\frac{\partial F_2}{\partial \omega}
%\end{array}\right|_{(c_0,\omega_0)}\notag\\
%=-4c_0q\omega_0 -4c_0\omega_0p^2 =-4c_0\omega_0(q+p^2)>0.
%\end{eqnarray}
\end{fmtext}

%%%%%%%%%%%%%%% End of first page %%%%%%%%%%%%%%%%%%%%%

\maketitle

\section{Introduction}
% When fluids are cooled down to near-zero temperatures, their properties undergo remarkable transformations. $^4{\rm He}$ is the only element that remains liquid close to zero temperature at atmospheric pressure. As liquid $^4{\rm He}$ is cooled below $T_{\lambda} = 2.17\ K$, it loses its viscosity and becomes a superfluid. The superfluidity in liquid $^4{\rm He}$ was first discussed by Kapitza~\cite{kapitza1938viscosity} and Allen and Misener~\cite{allen1938flow}. From the finite dissipation observed in the experiments of $^4{\rm He}$ at finite temperatures, Tisza~\cite{tisza1938transport} and Landau~\cite{landau1941j} suggested that at finite temperatures and below $T_{\lambda}$, liquid helium is a mixture of two fluids~\cite{balibar2017laszlo}: a normal component with finite viscosity and a superfluid component with zero viscosity. The densities of the two-component $\rho_n$ and $\rho_s$ are temperature dependent with the total density $\rho = \rho_n+\rho_s$. Soon after this, F. London~\cite{london1938lambda} suggested that the phenomenon of superfluidity is closely related to the Bose-Einstein condensation, which is a theory developed for non-interacting bosons by Bose and Einstein~\cite{Bose1924ZPhysik,Einstein1925}. The Bose-Einstein condensation is the process of accumulation of a large number of particles in the ground state below the transition temperature $T_C$, which was first realised in laboratory experiments in dilute atomic gases~\cite{Anderson_science1995}.

 Helium [$^4{\rm He}$] was first liquefied by Onnes~\cite{onnes1908condensation,Onnes1991}. About three decades after that Kapitza~\cite{kapitza1938viscosity} and Allen and Misener~\cite{allen1938flow} discovered that, when $^4{\rm He}$ is cooled below its $\lambda$-transition temperature $T_{\lambda} \simeq 2.17\ K$, it loses its viscosity and becomes a superfluid. 
 %From the finite dissipation observed in the experiments of $^4{\rm He}$ at finite temperatures, 
 Tisza~\cite{tisza1938transport} and Landau~\cite{landau1941j} suggested that, at temperatures $0 < T < T_{\lambda}$, liquid helium is a mixture of two fluids~\cite{balibar2017laszlo}: a finite-viscosity normal component and a zero-viscosity superfluid component. Soon after this, London~\cite{london1938lambda} hypothesized the intimate relation between superfluidity and Bose-Einstein condensation, whose theory, for \textit{non-interacting} bosons, was developed by Bose and Einstein~\cite{Bose1924ZPhysik,Einstein1925}. 
 
 The Gross-Pitaevskii (GP) equation~\cite{Gross_1961NCim,Pitaevskii_1961,dalfovo1999theory,pethick2008bose,Natalia_pnas2014} is used to model  Bose-Einstein condensates (BECs) in a system of weakly interacting bosons;
%\footnote{The GP equation is also known as the nonlinear Schr\"{o}dinger (NLS) equation.}; 
such a BEC was first realised in laboratory experiments in dilute atomic gases~\cite{Anderson_science1995}.
 % The GP equation describes a system of weakly interacting bosons at very low temperatures in which almost all the bosons accumulate in the lowest energy state.
 % The Bose-Einstein condensation is the process of accumulation of a large number of particles in the ground state below the transition temperature $T_C$, 
In terrestrial laboratories, superfluidity is found at extremely low temperatures (e.g., $T_{\lambda} \simeq 2.17 \ K$ for $^4{\rm He}$). 
%This naturally raises the question: could superfluidity exist beyond the confines of a lab, perhaps on a cosmic scale? 
% At first glance, it seems improbable due to the high temperatures and significant fluctuations typically found in celestial environments. However, certain regions of the universe do maintain remarkably low temperatures, though still not cold enough to sustain superfluidity. 

It was first pointed out by Migdal~\cite{MIGDAL1959655} that high-density regions, such as the interiors of neutron stars, could exhibit superfluidity of nuclear matter, with a transition temperature as high as $T_C\simeq 10^{11}\ K$ for densities $\rho\simeq 10^{14}\ g \cdot cm^{-3}$. Later Ginzburg and Kirzhnits~\cite{Ginzburg_1964} and  Ginzburg~\cite{Vitalii_L_Ginzburg_1969} suggested that the superfluidity of liquid $^4{\rm He}$ could occur at
$T_C \gtrsim 10^6\ K$ for densities $\rho \gtrsim 3 \times 10^7 \ g \cdot cm^{-3}$  on cosmic scales; and  Baym, Pethick, and Pines~\cite{Baym_1969Natur} proposed that charged particles, like the protons inside neutron stars, could also form a Type-II superconductor in such celestial bodies. Superfluidity and superconductivity in the cores of these dense objects lead to various dramatic effects, such as the drop in the surface temperatures of neutron stars~\cite{Pethic_RevModPhys_1992}, and provide a natural explanation for pulsar glitches~\cite{AK_Verma_PhysRevResearch.3.L022016,Shukla_PRD_2024}, which are sudden changes in the rotation frequency of a magnetised neutron star. At scales even larger than those of neutron stars ($\simeq 10 \ km$), superfluidity and BEC have been conjectured to be of relevance on galactic scales ($\simeq 1\ kpc$), e.g., for dark-matter haloes (DMH)~\cite{Sin_PhysRevD_1994,Ji_PhysRevD_1994,Guzman_1999}. 

Observations of the rotation curves of spiral galaxies~\cite{Rubin_1980ApJ,Rubin_1982ApJ} have provided clear evidence for some invisible dark matter. This dark matter forms almost spherical haloes around galaxies called dark matter haloes (DMH). The nature of dark matter particles that make up the DMH has been the topic of a long debate, but still, there is no unambiguous candidate.  It has been suggested that the dark matter particles can be bosons. The idea was first considered by Sin~\cite{Sin_PhysRevD_1994} and Ji and Sin~\cite{Ji_PhysRevD_1994} and independently by Guzman et al.~\cite{Guzman_1999} to model bosons as dark matter candidates, which can condense and form giant coherent matter waves around galaxies. For the formation of DMH around galaxies, the mass of these bosons should be of the order of $m\sim 10^{-24} eV/c^2$. The dark matter formed by these bosons has been given different names such as BEC dark matter (BECDM), fuzzy dark matter (FDM), and scalar-field dark matter (SFDM).

A natural description of the formation of Bose-Einstein condensate (BEC) is given by the Gross-Pitaevskii (GP) equation~\cite{Gross_1961NCim,Pitaevskii_1961}, which is also known as the nonlinear Schr\"{o}dinger (NLS) equation. The GP equation describes a system of weakly interacting bosons at very low temperatures in which almost all the bosons accumulate in the lowest energy state. This macroscopic accumulation of bosons can be described by a complex macroscopic wavefunction. Fluids around galaxies and inside neutron stars are strongly affected by their self-gravitational field. The central idea in describing self-gravitating Bose fluids around galaxies and the interior of neutron stars is to couple the GP equation with the gravitational field. The GP is the non-relativistic description of a system of bosons; hence, it is coupled with the Poisson equation for the gravitational potential, and the resulting equation is known as the Gross-Pitaevskii-Poisson equation (GPPE).

We present an overview of the GPPE framework and the real-time Ginzburg-Landau equations that have proved invaluable for theoretical investigations of gravitationally collapsed BECs  on galactic scales and of the interiors of dense celestial objects like neutron stars~\cite{Drummond_2017,Drummond_2017_b,AK_verma_2022,AK_Verma_PhysRevResearch.3.L022016,Verma_PhysRevResearch.4.013026,Shukla_PhysRevD.109.063009,Shukla_PRD_2024,Shukla_axion_PhysRevD.109.063009,PKM_PhysRevD.111.083511,sivakumar2025anomalousenergyinjectionturbulent}.

\section{Bosons and Axions: Dark Matter candidates}
\label{sec:BA}

Dark matter plays a significant role in our understanding of modern cosmology. It has a fascinating history going back to the dark bodies discussed by Kelvin~\cite{kelvin1904baltimore}, who gave a dynamical estimate for the amount of dark matter in the Milky Way. Inspired by this idea, Poincar\'e suggested the presence of \textit{mati\`ere obscure} around the Milky Way, referring to the obscure stars that move in galaxies with their existence unknown~\cite{hp1906pa}. The first concrete discussion of the existence of dark matter came from galaxy clusters and was presented in 1933 by Zwicky~\cite{zwicky6redshift,zwicky2009republication}, who studied the velocity dispersions of galaxies in the Coma cluster and used the virial theorem to estimate the mass of the cluster. From the number of observed galaxies and the average mass of a galaxy, Zwicky found a large discrepancy between the observed and calculated velocity dispersion, which gave a hint that dark matter is much greater in abundance than luminous matter. 

The first convincing observational evidence for dark matter came from the rotation curves of galaxies obtained by Rubin and Ford~\cite{rubin1970rotation}. The rotation curve of a galaxy is the circular velocity profile of stars and gas as a function of the distance from the galactic centre.  The circular velocity of different stars in the Milky Way can be determined by balancing the gravitational force with the centripetal force
\begin{eqnarray}
    \frac{m v_{\rm circ}^2}{r}=\frac{GMm}{r^2}\,; \quad v_{\rm circ} = \sqrt{\frac{GM}{r}}\,;
    \label{eq:circular_velocity}
\end{eqnarray}
%which is shown here 
[see \href{https://en.wikipedia.org/wiki/Galaxy_rotation_curve}{https://en.wikipedia.org} for an illustration].
%\begin{figure}[!hbt]
%    \centering
%    \includegraphics[scale=0.18]{figures/Rotation_curve_of_spiral_galaxy_Messier_33_(Triangulum).png}
%    \caption{Rotation curve of stars in the Milky Way. The dashed curve represents the expected rotational velocity based on visible matter alone, while the solid curve accounts for the presence of dark matter [\href{https://en.wikipedia.org/wiki/Galaxy_rotation_curve}{https://en.wikipedia.org}]. }
%    \label{fig:rotation_curve}
%\end{figure}
This suggests that, as we move away from the galactic centre, the speed of stars should decrease. However, this is contrary to observations that yield rotation curves in which the velocities of stars remain constant, or even increase, with distance from the galactic center; this indicates the presence of dark matter. 
% as shown by the solid curve in [\href{https://en.wikipedia.org/wiki/Galaxy_rotation_curve}{https://en.wikipedia.org}]. The observed 
Following the observational evidence of dark matter from galaxy rotation curves, the study of dark matter has become a central topic in modern cosmology \cite{Bertone_RevModPhys}. This is particularly relevant in the context of the $\Lambda$-cold-dark-matter ($\Lambda$CDM) model, where $\Lambda$ represents the cosmological constant. 

%After gaining some observational evidence, astronomers started thinking about 
To understand the constituents of dark matter, astronomers initially hypothesized that these objects were similar to ordinary stars but with much lower luminosity, coining the term MACHOs (massive astrophysical compact halo objects), which include planets, white dwarfs, neutron stars, red dwarfs, and black holes. However, high-precision measurements of the cosmological baryon density~\cite{Burles_1998} showed that baryonic matter constitutes less than $20\%$ of the universe's total matter, leaving little room for MACHOs as the primary component of dark matter.~\footnote{Milgrom proposed that, instead of introducing a new dark matter candidate, we should modify Newtonian dynamics, a theory which became known as modified Newtonian dynamics (MOND)~\cite{Milgrom_1983a, Milgrom_1983b,khoury2022dark}. In MOND, rather than using Newton's second law $F=ma$, the acceleration is scaled as $ F = m\frac{a^2}{a_0}$, where $a$ varies slowly for $a\ll a_0\sim1.2\times 10^{-10}\ {\rm m\cdot s^{-2}}$. This modification accounted for the observed acceleration of stars in the rotation curve at distances far from the galactic center. However, it later became clear that, although MOND works well on the scale of individual galaxies, it is unsuccessful at explaining phenomena at much larger scales, such as those observed in galaxy clusters.}

Attention has now shifted towards unravelling the particle nature of dark matter. Neutrinos, which are stable particles from the standard model with weak electromagnetic and strong interactions, have been considered possible candidates for dark matter. Particles with these characteristics are collectively known as WIMPs (weakly interacting massive particles)~\cite{Abdallah_PhysRevLett.120.201101}. Although there have been several comprehensive searches for WIMPs, they have not succeeded in obtaining unambiguous signatures for such particles~\cite{Abdallah_PhysRevLett.120.201101,Aprile_PhysRevLett.121.111302}. A search for boosted dark matter interacting with electrons has also proved to be infructuous~\cite{kachulis2018search}.
%do not adequately account for the large-scale structure of the universe.

%After gaining some understanding of the particle nature of dark matter, a model is needed to study its dynamics. 
In cosmology, the $\Lambda$-CDM model incorporates WIMPs, describing them as a system of collisionless particles. This $\Lambda$-CDM model works well at large (cosmological) distances, but it encounters problems at small (galactic) scales. These problems include the core-cusp problem \cite{Moore_MNRAS_1999} and the missing-satellite problem \cite{Kamionkowski_PhysRevLett.84.4525}.
%with the core-cusp problem being particularly well-known.

%In $N$-body simulations using the $\Lambda$-CDM model, the density profile at the center of a galaxy shows a cusp, which contradicts experimental observations suggesting a flat core density. The core-cusp problem is illustrated schematically in Fig.~\ref{fig:core_cusp}(a), with actual observations shown in Fig.~\ref{fig:core_cusp}(b).
%\begin{figure}[!hbt]
%    \centering
%    \includegraphics[scale=0.2]{figures/core_cusp.jpeg}
%    \caption{The variation of dark matter density as a function of distance from the galactic center showing core-cusp problem of the $\Lambda$-CDM model: {\bf (a)} a schematic, and {\bf (b)} based on real observational data (taken from Ref.~\cite{Oh_2011}).}
 %   \label{fig:core_cusp}
%\end{figure}

%While these discussions about the particle nature of dark matter continue, 
Another radical idea, which invokes the wave nature of dark-matter particles, can also cure the core-cusp problem~\cite{Oh_2011} and other small-scale crises of the $\Lambda$-CDM model. This considers the quantum nature of dark matter particles, which manifests itself when the mass of the dark matter particle is so small that their de Broglie wavelengths are of the order of 
dark-matter halos, as illustrated schematically in Fig.~\ref{fig:Dark_matter_halo}(a). In this \textit{ultra-light dark matter} (ULDM), the wavelengths of these particles overlap significantly, allowing them to form a Bose-Einstein condensate (BEC), which gives a coherent and flat density profile near the galactic centre. The self-interaction between the condensed particles in this BEC also affects the distribution of dark matter. This self-interaction is repulsive,  with a positive scattering length, for conventional bosons; by contrast, attractive self-interactions, with negative scattering length, occur for axions~\cite{Braaten_RevModPhys.91.041002,ringwald2024review}.

\subsection{Bosons}
\label{sec:boson_stars}

Despite numerous predictions and experimental searches, there is still no unambiguous candidate for dark matter \cite{Rott:2017mxp, Aprile_PhysRevLett.121.111302}. While these searches continue, it is worthwhile to explore the self-gravitating systems of ULDMs or BECs discussed above. Regarding the large-scale structure formation after the Big Bang, it has been suggested that an ultra-light pseudo-Goldstone boson could be a candidate for dark matter to explain the "\textit{late-time cosmological phase transition}"~\cite{Hill_1988vm}. It was also inferred that, to explain the large-scale structures of the universe, the Compton wavelength of these bosons must be of the order of these structures, so they must be extremely light.
\begin{figure}[!hbt]
    \centering
    \includegraphics[scale=0.18]{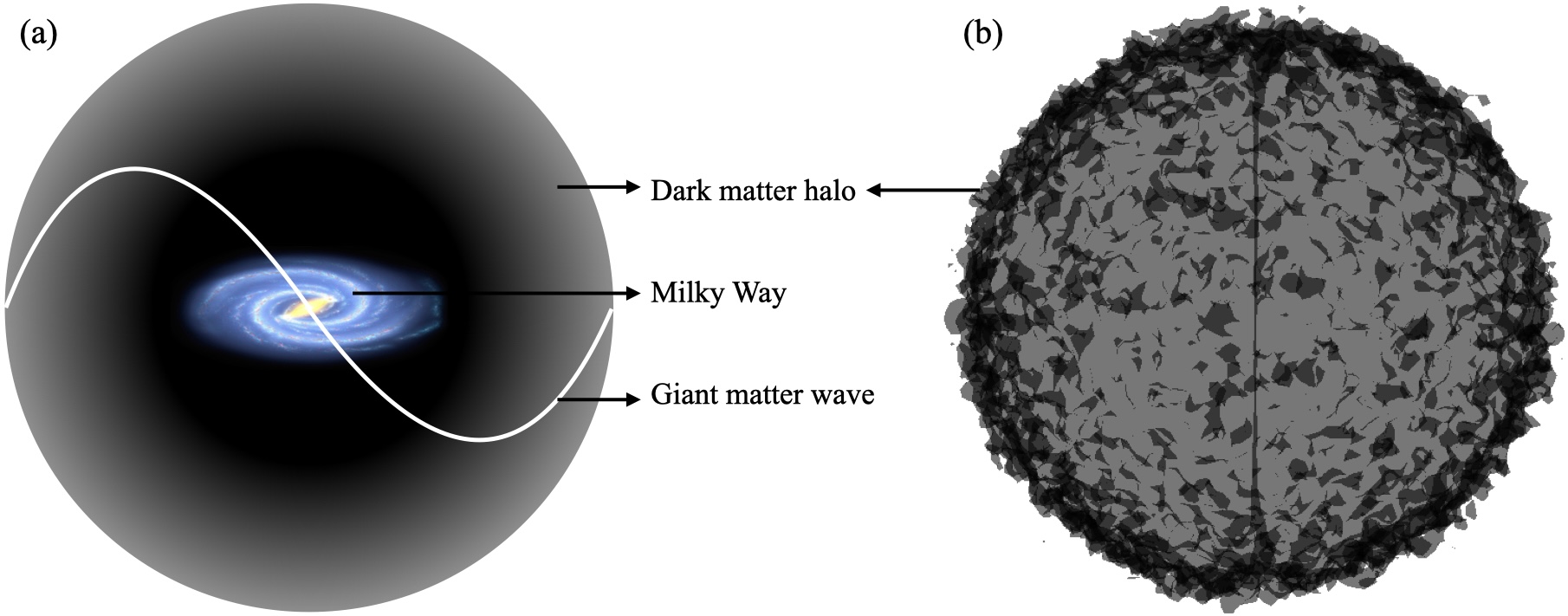}
    \caption{{\bf (a)} A schematic diagram of the spherical distribution of the dark-matter halo (say around the Milky Way) composed of ultra-light bosons that exhibit wave-like properties. {\bf (b)} Dark-matter halo from our direct numerical simulation (DNS) of the Gross-Pitaevskii-Poisson equation~\eqref{eq:GP_poisson}.}
    \label{fig:Dark_matter_halo}
\end{figure}

Figure \ref{fig:Dark_matter_halo}(a) provides a schematic illustration of the spherical distribution of dark matter around a galaxy (say the Milky Way). The mass of a bosonic dark matter particle can be estimated using $\lambda = \frac{h}{mv}$.
%\begin{eqnarray}
%    \lambda = \frac{h}{mv}\,.
%    \label{eq:halo_wavelength}
%\end{eqnarray}
If the bosons condense and form a giant matter wave, with a wavelength of the order of the halo [Fig.~\ref{fig:Dark_matter_halo} (a)] around the galaxy (i.e., $\lambda\sim 10$ kpc), then $m\sim 10^{-22}\ {\rm eV/c^2}$, with $v$, the average saturation speed, determined using the rotation curve [$v\simeq220 \ km/s$].

These ultra-light bosons condense to form a Bose-Einstein condensate on a galactic scale and can be described by a macroscopic complex wavefunction $\psi$. In the absence of self-interaction between the bosons, the wavefunction $\psi$ satisfies the Schr\"{o}dinger-Poisson equation~\cite{Sin_PhysRevD_1994,Ji_PhysRevD_1994}:
\begin{eqnarray}
    i\hbar\frac{\partial \psi}{\partial t} = -\frac{\hbar^2}{2m}\nabla^2 \psi +m \Phi \psi\,; \quad \nabla^2\Phi = 4\pi Gm|\psi|^2\,;   \label{eq:Schrodinger_poisson}
\end{eqnarray}
here, $\Phi$ is the Newtonian gravitational potential
%\begin{eqnarray}
%    \nabla^2\Phi = 4\pi Gm|\psi|^2.
%    \label{eq:poisson_eq}
%\end{eqnarray}
and $G$ is Newton's gravitational constant. This non-relativistic treatment of a system of bosons of mass $m$ is applicable under certain limits: For a system of size $R$ and mass $M$, the circular velocity of the dark-matter mass distribution is  $v^2= GM/R$ [Eq.~\eqref{eq:circular_velocity}]. We can determine the length scale at which the quantum pressure term becomes significant by balancing the two terms on the right-hand side of Eq.~\eqref{eq:Schrodinger_poisson}, which yields $r_0 =\tfrac{\hbar^2}{m^2GM}$. Therefore, for $R\gg r_0$, we require~\cite{Sin_PhysRevD_1994} $ v^2 \ll (GMm)^2$ (with $\hbar=c=1$).
%\begin{eqnarray}
%    v^2 \ll (GMm)^2\,.
%\end{eqnarray}
This Newtonian limit is valid when $GMm\ll 1$; for the mass scales $M\sim 10^{12}M_{\circ}$ and $m\sim 10^{-24}\ {\rm eV}$ this yields $v\ll 10^{-2}$.

If there are repulsive interactions between the bosons constituting dark matter, the wavefunction $\psi$ satisfies the Gross-Pitaevskii-Poisson equation (GPPE)
\begin{eqnarray}
    i\hbar\frac{\partial \psi}{\partial t} = -\frac{\hbar^2}{2m}\nabla^2 \psi +g|\psi|^2\psi+m \Phi \psi\,;\quad \nabla^2\Phi = 4\pi G(m|\psi|^2-\rho_{\rm bg})\,;
    \label{eq:GP_poisson}
\end{eqnarray}
where $g=4\pi a_s\hbar^2/m$ is the interaction strength, $a_s>0$ is the scattering length, and $\Phi$ is the gravitational potential that satisfies the Poisson equation~\eqref{eq:Schrodinger_poisson}. The subtraction of the mean background density is often called the Jeans swindle~\cite{KIESSLING_2003}. The Schr\"{o}dinger-Poisson system of Eqs.~\eqref{eq:Schrodinger_poisson}, with no self-interaction, has been used extensively to study dark-matter haloes around different galaxies~\cite{Wayne_PhysRevLett_2000,Bernal_PhysRevD_2006,Guzman_PhysRevD_2003,Guzman_PhysRevD_2004}. In the case of self-interaction between bosons, the GPPE model has been used likewise to explore different aspects of self-gravitating bosonic systems, such as the mass-radius relations for boson stars using a Gaussian Ansatz~\cite{Chavanis_2011,Chavanis_2016}.

The self-gravitating bosons, described by the GPPE~\eqref {eq:GP_poisson}, collapses under its own gravity
for wavenumbers $k$ below $ k_J$, the Jeans wavenumber. The linearisation of Eq.~\eqref {eq:GP_poisson}, about the constant density $|\psi|^2=n_0$, yields $k_J$ and the following dispersion relation for the frequency 
%$\omega$ and wavenumber $k$ with the Jeans wavenumber $k_J$ as
\begin{eqnarray}
    \omega(k) = \sqrt{\frac{\hbar^2k^4}{(2m)^2} -\frac{k^2}{m} \left(\frac{Gn_0}{k^2} -gn_0\right)}\,; \quad k_J^2 &=& \frac{2mgn_0}{\hbar^2} \bigg[-1+ \sqrt{1+ \frac{G\hbar^2 }{mg^2n_0}}
 \bigg]\,.
\label{eq:disprel}
\end{eqnarray}
%whence we define the wave number
%\begin{equation}
%\begin{aligned}
%    k_J^2 &=& \frac{2mgn_0}{\hbar^2} \bigg[-1+ \sqrt{1+ \frac{G\hbar^2 }{mg^2n_0}} 
% \bigg]\,.
 %\end{aligned}
%\end{equation}
The GPPE~\eqref{eq:GP_poisson} describes self-gravitating bosons at zero temperature $T=0$. To obtain $T > 0$ effects, we 
follow earlier GPE and GPPE~\cite{krstulovic2011dispersive,krstulovic2011energy,shukla2013turbulence,Natalia_pnas2014,AK_verma_2022,AK_Verma_PhysRevResearch.3.L022016} studies that employ the truncated version of the GPPE~\cite{Natalia_pnas2014}, projected onto a finite number of Fourier modes. We use the Fourier transform $\psi ({\bf x}) = \sum_{ {\bf k} } \hat{\psi}_{{\bf k}} \exp (i {\bf k} \cdot {\bf x})$ %follows
%\begin{eqnarray}
%\psi ({\bf x}) = \sum_{ {\bf k} } \hat{\psi}_{{\bf k}} \exp (i {\bf k} \cdot {\bf x}) \,, 
%\label{eq:Fourier}
%\end{eqnarray}
and truncate the modes as follows: $\hat{\psi} \equiv 0$ for $ |{\bf k}| > k_{max}$, with $k_{max} = [N/3]$, where $N^3$ is the number of collocation points. The Fourier-truncated GPPE (T-GPPE) is
\begin{eqnarray}
i\hbar \frac{\partial \psi}{\partial t} &=& P_G \bigg[-\frac{\hbar^2}{2m} \nabla^2 \psi + P_G \{ ( G \nabla^{-2} + g)  |\psi|^2 \} \psi\bigg] \,,
\label{eq:TGPPE}
\end{eqnarray}
with the Galerkin projector $ P_G[\hat{\psi}_{{\bf k}}] = \theta (k_{max} - |{\bf k}|) \hat{\psi}_{{\bf k}} $.

A system of self-gravitating bosons, evolving under the T-GPPE~\eqref{eq:TGPPE}, takes a long time to converge to the ground state because of finite-amplitude thermal fluctuations and sound waves. This convergence can be accelerated using the imaginary-time approach to study equilibrium properties. The imaginary-time ($t\to -it$) version Eq.~\eqref{eq:TGPPE} is the stochastic Ginzburg-Landau-Poisson (SGLP) equation (we omit $P_G$ for notational simplicity):
\begin{eqnarray}
     \hbar\frac{\partial \psi}{\partial t} = \frac{\hbar^2}{2m}\nabla^2 \psi -g|\psi|^2\psi-m \Phi \psi +\sqrt{\frac{2\hbar}{\beta}} \xi({\bf x},t)\,,
    \label{eq:SGLP_poisson}
\end{eqnarray}
where $\beta = 1/(k_B T)$, $k_B$ and $T$ are, respectively, the Boltzmann constant and the temperature; the Gaussian noise $\xi({\bf x},t)$ has  zero mean and the variance $\left\langle \xi({\bf x},t) \xi^*({\bf x'},t') \right\rangle = \delta(t -t') \delta({\bf x} - {\bf x'})$. The SGLP equation~\eqref{eq:SGLP_poisson} contains an explicit term for the control of the temperature [this is not possible in the T-GPPE~\eqref{eq:TGPPE}].

We solve the real-time TGPPE~\eqref{eq:TGPPE} and imaginary-time SGLPE~\eqref{eq:SGLP_poisson} using the pseudospectral method in a cubic domain of resolution $N^3$ with periodic boundary conditions in all three directions. Fig.~\ref{fig:Dark_matter_halo}(b) shows collapsed self-gravitating bosons using the real-time GPP~\eqref{eq:TGPPE} equation. Besides a spherical mass distribution representing a dark matter halo, we also observe many fluctuations which come from the finite temperature effects of the truncated GPPE. We also obtain the formation of speherical haloes using the imaginary-time SGLP~\eqref{eq:SGLP_poisson} equation. 
\begin{figure}
    \centering
    \includegraphics[scale=0.3]{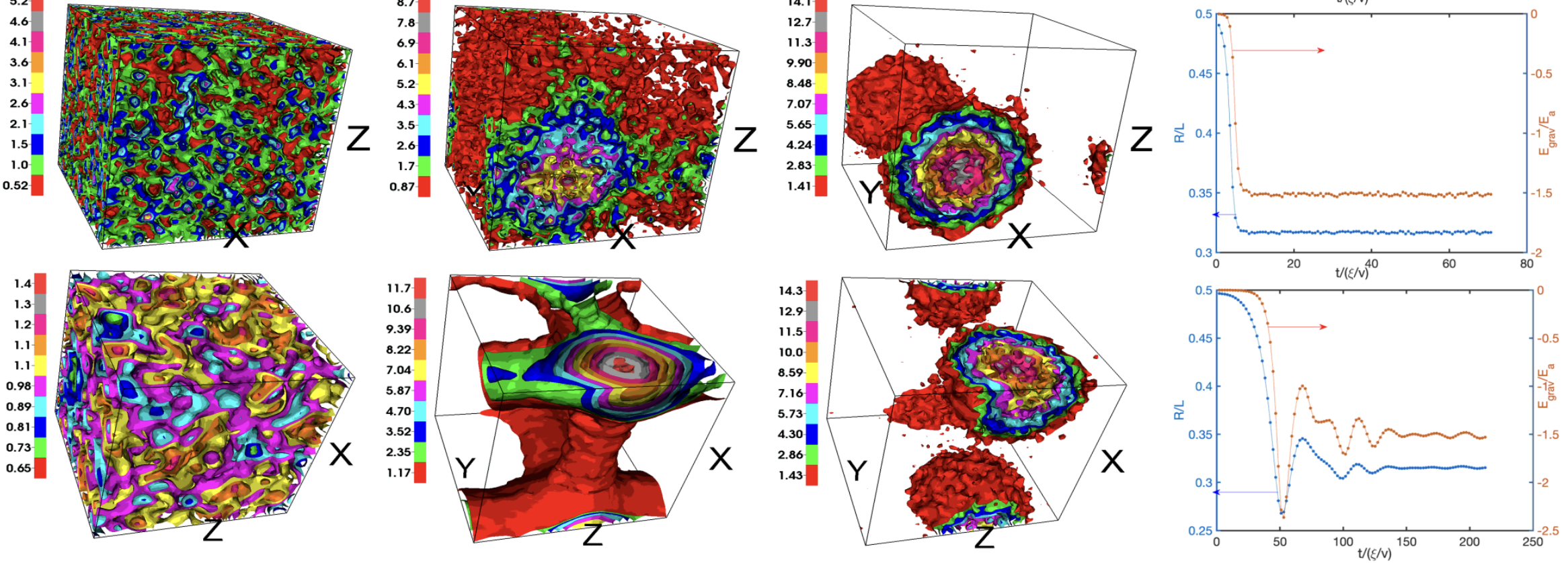}
    \caption{Columns 1–3 show 10-level contour plots of density $|\psi|^2$ at representative times using SGLPE (imaginary-time) in the top row and GPPE (real-time) in the bottom row, respectively. Column 4: Plots of the scaled radius of gyration $R/L$ [\eqref{eq:radius_gyration}] versus the scaled time [adapted from Ref.~\cite{AK_Verma_PhysRevResearch.3.L022016} with permission from the APS].}
    \label{fig:Bosons_AKV}
\end{figure}
Figures~\ref{fig:Bosons_AKV} show the formation of collapsed bosons by comparing real-time GPPE and imaginary-time SGLPE. Row1 and Row2 of Figs.~\ref{fig:Bosons_AKV} show collapsed bosonic condensates using imaginary-time SGLPE~\eqref{eq:SGLP_poisson} and real-time GPPE~\eqref{eq:TGPPE} at temperature $T\neq 0$, respectively. Both evolutions show similar thermal fluctuations. To show a quantitative agreement of steady ground state solutions of real-time TGPPE and imaginary-time SGLPE, column4 of  Figs.~\ref{fig:Bosons_AKV} shows the plots of radius of gyration and the gravitational energy defined as follows
\begin{eqnarray}
    \frac{R}{L} = \frac{1}{L}\sqrt{\frac{\int \rho(r)r^2 d{\bf r}}{\int \rho(r) d{\bf r}}}\,; \quad  E_{\rm grav} = \frac{G}{2}\int d^3 x |\psi|^2 \nabla^{-2}|\psi|^2\,.
    \label{eq:radius_gyration}
\end{eqnarray}
where $L$ is the length of the simulation domain. 

\subsection{Axions}

Another class of bosons, called axions~\cite{Braaten_RevModPhys.91.041002,ringwald2024review,semertzidis2022axion}, are also promising candidates for dark matter. They arose naturally in the context of the strong-CP problem of quantum chromodynamics (QCD)~\cite{Kim_RevModPhys.82.557}, a solution for which was proposed by Peccei and Quinn \cite{Peccei_PhysRevD.16.1791} by introducing a global $U(1)$ symmetry; a spontaneous breaking of this symmetry led a Nambu-Goldstone boson~\cite{weinberg1978the_axion,wilczek1978problem}, now known as the axion. 
In QCD, the mass range for axions is $m_a\sim 10^{-6}-10^{-4}$ eV/$\rm c^2$; such QCD axions have an extremely weak self-interaction. 

%It is important to note that when comparing 
There is a significant difference between the mass of QCD axions [$m_a\sim 10^{-6}-10^{-4}$ eV/$\rm c^2$] and the dark-matter bosonic particles [$m\sim 10^{-24}\ $ eV/$\rm c^2$] discussed in Section~\ref{sec:boson_stars}, . Therefore, QCD axions cannot form galactic-scale halos as Bose-Einstein Condensate (BEC) dark matter does. Instead, they form mini axion stars that constitute dark matter halos as mini compact objects~\cite{Chavanis_2018}. There are other types of axions with very small masses, known as ultra-light axions (ULA)~\cite{MARSH20161}. These ULAs have masses of the order of $m\sim 10^{-22}\ {\rm eV}$ and can form large structures like dark-matter halos around galaxies.

Axions could have been produced in the early universe, e.g., by thermal mechanisms~\cite{Kephart_PhysRevLett.58.171} and vacuum misalignment~\cite{PRESKILL1983127}. Axions have a large occupation number so they are extremely non-relativistic and they can be described by a classical field $\psi$. 
In the non-relativistic limit, this classical field satisfies the GPPE~\eqref{eq:GP_poisson}, but with the key difference that axions have a negative scattering length ($a_s<0$), i.e., an attractive self-interaction~\cite{Braaten_RevModPhys.91.041002} that leads to an instability which can be controlled by higher-order terms in the axion potential (see below). The possibility of a gravitationally bound axionic condensates, or \textit{axion stars},  was first hypothesized by Tkachev~\cite{Tkachev_1986SvAL...12..305T}.

\begin{figure}
    \centering
    \includegraphics[scale=0.7]{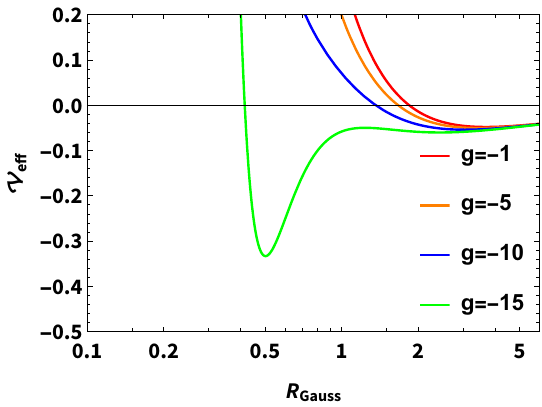}
    \put(-150,120){{ (a)}}
    \put(-35,89){{\tiny \bf LM}}
    \put(-90,50){{\tiny \bf GM}}
    \includegraphics[scale=0.7]{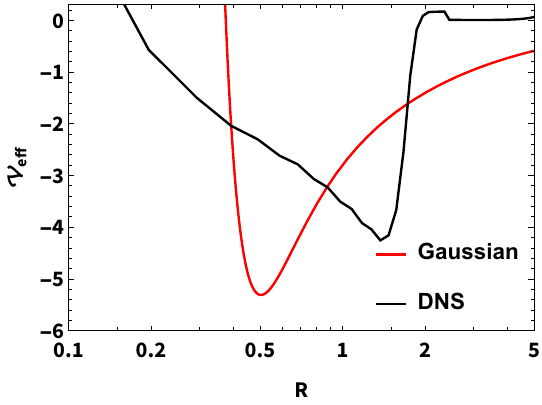}
    \put(-150,120){{ (b)}}
    \caption{(a) Log-linear plots of the effective potential~\eqref{eq:eff_pot} versus the axionic condensate’s radius. The minima labeled ${\bf LM}$ and ${\bf GM}$ correspond, respectively, to low- and high-density phases. (b) Plots of Eq.~\eqref{eq:eff_pot} versus the radius of gyration obtained by using the Gaussian Ansatz (red) and our DNS (black) [adapted from Ref.~\cite{Shukla_PhysRevD.109.063009}, with permission from the APS].}
    \label{fig:effective_potential}
\end{figure}

If axions are described by a scalar field $\phi$, the Lagrangian can be written as:
\begin{eqnarray}
    \mathcal{L} = \frac{1}{2}\partial_{\mu}\phi \partial^{\mu}\phi- V(\phi)\,; \quad  V(\phi) = \frac{m_a^2f_a^2c}{\hbar^3}\bigg[1-\cos\bigg(\frac{\sqrt{\hbar c}\phi}{f_a}\bigg)\bigg]- \frac{c^2}{2\hbar^2}m_a^2\phi^2\,,
    \label{eq:axion_lag}
\end{eqnarray}
where $V(\phi)$ is the widely used instanton potential in most phenomenological theories of axions, 
%\begin{eqnarray}
%    V(\phi) = \frac{m_a^2f_a^2c}{\hbar^3}\bigg[1-\cos\bigg(\frac{\sqrt{\hbar c}\phi}{f_a}\bigg)\bigg]- \frac{c^2}{2\hbar^2}m_a^2\phi^2\,,
%    \label{eq:instanton_potential}
%\end{eqnarray} 
$m_a$ is the mass of the axion and $f_a$ is the decay constant. The last term in Eq.~\eqref{eq:axion_lag} for $V(\phi)$ is the mass term,  which is subtracted to highlight the interaction terms. Expanding the cosine term in a Taylor series~\cite{Chavanis_2018}
\begin{eqnarray}
    V(\phi) =-\frac{m_a^2f_a^2c}{\hbar^3}\sum_{n=2}^{\infty}\frac{(-1)^n}{(2n)!}\bigg(\frac{\sqrt{\hbar c}\phi}{f_a}\bigg)^{2n}\,.
\end{eqnarray}
To convert the real scalar field $\phi$ into the complex scalar $\psi$, we use the transformation~\cite{Chavanis_2018}
\begin{eqnarray}
    \phi^{2n} = \frac{1}{2^n}\bigg(\frac{\hbar}{m}\bigg)^{2n} \frac{(2n)!}{(n!)^2}|\psi|^{2n}\,,
    \label{eq:transformation}
\end{eqnarray}
which gives the potential
\begin{eqnarray}
     V(|\psi|^2) =-\frac{m_a^2f_a^2c}{\hbar^3}\sum_{n=2}^{\infty}\frac{(-1)^n}{(n!)^2}\bigg(\frac{\hbar^3 c |\psi|^2}{2f_a^2 m_a^2}\bigg)^{n}\,.
     \label{eq:axion_general_potential}
\end{eqnarray}
For $n=2$, Eq.~\eqref{eq:axion_general_potential} simplifies to
\begin{eqnarray}
    V = -\frac{\hbar^3c^3}{16f_a^2m_a^2}|\psi|^4\,,
    \label{eq:quartic_potential}
\end{eqnarray}
 the quartic potential, used in the GPPE~\eqref{eq:GP_poisson} for bosonic dark matter \textit{but with a negative sign}. The potential~\eqref{eq:quartic_potential} has been used to study the mass-radius relation of a self-gravitating system with attractive self-interaction; not surprisingly, the system is extremely unstable above a very low critical mass~\cite{Chavanis_2016}, because the repulsive quantum force cannot counterbalance the combined effects of the attractive gravitational and self-interaction forces. To stabilize this self-gravitating axionic condensate with attractive self-interaction, we need to consider the next term, $n=3$, in the series expansion of Eq.~\eqref{eq:axion_general_potential}, namely,
 \begin{eqnarray}
      V = -\frac{\hbar^3c^3}{16f_a^2m_a^2}|\psi|^4 + \frac{\hbar^6c^4}{288m_a^4 f_a^4}|\psi|^6\,.
      \label{eq:cubic_quintic_potential}
 \end{eqnarray}
This potential~\eqref{eq:cubic_quintic_potential} yields the cubic-quintic GPPE, which has been used to study the dense and dilute phases of axion stars using a Gaussian Ansatz at temperature $T=0$~\cite{Chavanis_2018}. Specifically, if we use the potential~\eqref{eq:cubic_quintic_potential} in the axionic Lagrangian~\eqref{eq:axion_lag} and make the replacement~\eqref{eq:transformation}, we obtain the following cubic-quintic Gross-Pitaevskii-Poisson equation (cq-GPPE):
\begin{eqnarray}
    i\hbar\frac{\partial \psi}{\partial t} = -\frac{\hbar^2}{2m}\nabla^2 \psi +g|\psi|^2\psi+g_2|\psi|^4\psi+m \Phi \psi\,,
    \label{eq:cqGPP_poisson}
\end{eqnarray}
where $g<0$ is the coefficient of the cubic term describing attractive self-interaction, and $g_2>0$ is the coefficient of the quintic stabilising term. For the axion potentail~\eqref{eq:cubic_quintic_potential}, these coefficients are
\begin{eqnarray}
    g=-4\pi a_s\hbar^2/m\,; \quad g_2= \frac{32\hbar^4 \pi^2 a^2}{3m^3 c^2}\,,
\end{eqnarray}
where $a_s$ is the scattering length, $m$ is the mass of axion, and $c$ is the speed of light. We project the cq-GPPE onto a finite number of Fourier modes to include finite temperature effects as we had done earlier for
the cubic case, and obtain the following truncated cq-TGPPE:  
\begin{eqnarray}
i\hbar \frac{\partial \psi}{\partial t} &=& P_G \bigg[-\frac{\hbar^2}{2m} \nabla^2 \psi + P_G \biggl\{ \bigg( G \nabla^{-2} + g + g_2 P_G (|\psi|^2)\bigg)  |\psi|^2 \biggr\} \psi\bigg] \,.
\label{eq:cq_TGPPE}
\end{eqnarray}
To study the equilibrium configurations of axion stars, we solve the imaginary-time version ($t\to -it)$ of the cq-TGGP equation. Equation~\eqref{eq:cq_TGPPE} comes from a Hamiltonian with two conserved quantities, namely, the number of particles $N=\int |\psi|^2 d^3x$, and the total energy $E = E_k + E_{int}+E_G$, where $E_k$, $E_{int}$, and $E_G$ are the following kinetic, self-interaction, and gravitational energies, respectively:
\begin{eqnarray}
E_{k}  =  \frac{\hbar^2}{2m} \int d^3 {\bf x} |\nabla \psi |^2 \,;\;\;
%\nonumber \\
E_{int} &=&  \int d^3 {\bf x} \left[  \frac{g}{2} (P_G |\psi|^2)^2 + \frac{g_2}{3} \left[  P_G \left\lbrace (P_G |\psi|^2)^2\right\rbrace \right] |\psi|^2  \right] \,; \nonumber \\
E_G &=& \frac{G}{2} \int d^3 {\bf x} \left[ P_G |\psi|^2 \right] \nabla^{-2}  \left[ P_G |\psi|^2 \right] \,.
\label{eq:energy}
\end{eqnarray} 
We use the Madelung transform $\psi({\bf r},t) = \sqrt{\tfrac{\rho}{m}} e^{i\vartheta({\bf r},t)}$ to write the kinetic energy as $E_k = E_{kc} + E_{q}$, which is the sum of the classical and quantum kinetic energies:
\begin{eqnarray}
E_{kc} = \frac{1}{2} \int \rho {\bf v}_s^2d^3{\bf x}\,; \quad {\bf v}_s = \frac{\hbar}{m} \nabla \vartheta\,; \quad E_{kq} = \frac{\hbar^2}{8m^2} \int \frac{(\nabla \rho)^2}{\rho}d^3{\bf x}\,.
\end{eqnarray}
%where ${\bf v}_s = \frac{\hbar}{m} \nabla \vartheta$. 
The total energy can now be written as $E = E_{kc} + \mathcal{V}_{eff}$, with the effective potential
\begin{eqnarray}
    \mathcal{V}_{eff} = E_{kq}  +E_{int}+E_G\,.
    \label{eq:eff_pot}
\end{eqnarray}
The stable configurations of axion stars described by the cq-TGPP~\eqref{eq:cq_TGPPE} are different from those governed by the TGGP~\eqref{eq:TGPPE} for bosons. This difference arises because of the different form of the self-interaction potential of the axion~\eqref{eq:axion_general_potential}, which contains two nonlinear terms. 
To understand equilibrium configurations, we use the following Gaussian Ansatz for a spherical density distribution:
\begin{eqnarray}
\rho(r) = \rho(0) e^{-r^2/R_{Gauss}^2} \,,
\label{eq:gaussian_ansatz}
\end{eqnarray}
where $\rho(0) = M/(\pi^{3/2} R_{Gauss}^3)$, and $R_{Gauss}$ sets the scale of the radius of the axion star. Figure~\ref{fig:effective_potential}(a) shows $\mathcal{V}_{eff}$ as a function of $R_{Gauss}$ [using the Ansatz~\eqref{eq:gaussian_ansatz}], for different values of the parameters. There is a low-density minimum at small negative values of $g$. A new minimum appears as $g$ becomes more negative, so there are two minima [${\bf LM}$ and ${\bf GM}$ for low- and high-density phases of the axionic system]. We refer the reader to Ref.~\cite{Shukla_axion_PhysRevD.109.063009} for details. From our DNS we obtain $\mathcal{V}_{eff}$ and compare it with its
Gaussian-Ansatz~\eqref{eq:eff_pot} counterpart [see Fig.~\ref{fig:effective_potential}(b)].
% , we present plots of $\mathcal{V}_{eff}$ versus the radius of gyration $R$ from our DNS (in black) and the Gaussian Ansatz (in red), for $g=-15$, $g_2=8$, $G=100$, and $M=0.97$. 
This Ansatz does not yield quantitatively accurate positions of minima,
%of $\mathcal{V}_{eff}$, for $g=-15$, $g_2=8$
but it does show a stable axionic condensate~\footnote{To compare our DNS results with their Gaussian-Ansatz counterparts, we must account for the following factor: $R_{Gauss} = \sqrt{2/3} R$, with $R$ the radius of gyration~\eqref{eq:radius_gyration}).}.

\begin{figure}
    \centering
    \includegraphics[scale=0.25]{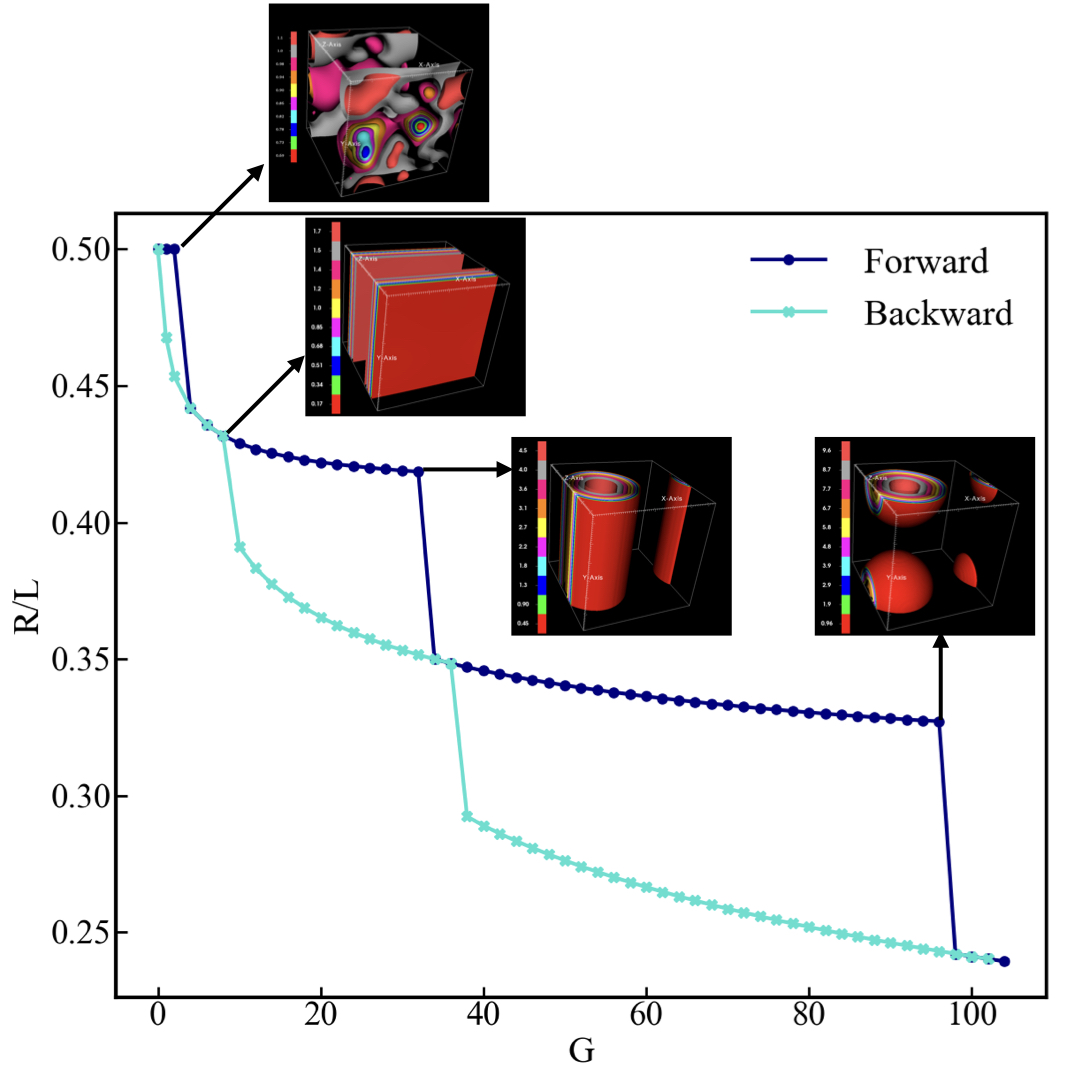}
    \caption{Plots of the scaled radius of gyration $R/L$ versus the gravitational interaction parameter $G$ in the cq-GPPE; blue and green curves show, respectively, curves along which $G$ increases and decreases [adapted from Ref.~\cite{Shukla_PhysRevD.109.063009}, with permission from the APS].}
    \label{fig:rad_vs_G}
\end{figure}

% Our Gaussian-Ansatz study of $\mathcal{V}_{eff}$ shows that, for $g=-15$ and $g_2=8$, a stable axionic condensate may occur. 
We can investigate the formation of structures in such axionic condensates by varying the strength $G$ of gravity: We start from a uniform density distribution state with small-amplitude perturbations and numerically integrate the imaginary-time version of Eq.~\eqref{eq:cq_TGPPE} for a small value of $G$. After obtaining the equilibrium state, we calculate the radius of gyration $R$ using Eq.~\eqref{eq:radius_gyration}. We continue this analysis with increasing values of $G$ and calculate the radius of gyration. Figure~\ref{fig:rad_vs_G} shows the normalised radius of gyration $R/L$ as a function of $G$. With increasing $G$, the system displays three first-order transitions: (a) from a statistically homogeneous state to pancakes; (b) next to a state with a cylindrical distribution; and (c) finally to a spherical assembly. $R/L$ jumps at these transitions. Metastability makes these jumps appear as hysteresis loops~\cite{rao1990magnetic}, different branches of which appear with increasing-$G$ (blue) and decreasing-$G$ (green) scans.

%We obtain the spherically collapsed equilibrium states of axions in a non-rotating frame. 
We show below how the axion condensate is modified in the presence of rotation,
arising from tidal torques on the axionic assembly from its surroundings. 
%One of the most remarkable features of superfluid axions, 
When we rotate the axion star with an angular
frequency $\Omega$, we obtain quantized vortices, if $\Omega > \Omega_c $, a critical angular frequency. The circulation around the  vortex line is quantized as $\oint_C {\bf v}_s \cdot {\bf dl} = n\kappa$,
%\begin{eqnarray}
%\oint_C {\bf v}_s \cdot {\bf dl} = n\kappa \,,
%\label{eq:circulation}
%\end{eqnarray}
where ${\bf v}_s$ is the superfluid velocity, $n$ is an integer, and $\kappa \equiv h/m$. We investigate the formation of such vortices by introducing the rotation term $-\Omega L_z\psi$ into the cq-GPPE~(\ref{eq:cqGPP_poisson}),
 where $L_z = -i\hbar (x\partial_y - y \partial_x)$ is the $z$-component of the angular momentum $\bf L = x\times P$. The equilibrium configuration can then be obtained 
 by using the following imaginary-time version with the rotation term:
\begin{eqnarray}
\hbar \frac{\partial \psi}{\partial t} &=& \frac{\hbar^2}{2m} \nabla^2 \psi - \big[ G \Phi + g |\psi|^2 + g_2 |\psi|^4 - \Omega L_z\big] \psi \,. 
\label{eq:cq_SGLPE_rot}
\end{eqnarray}
To study the rotational dynamics of a collapsed axion star, we start with an initial condition with a uniform density distribution and superimposed small-amplitude perturbations in Eq.~(\ref{eq:cq_SGLPE_rot}), with $\Omega=0$; for the chosen set of parameters, this yields a spherical collapsed star, which we then use as the initial condition for Eq.~(\ref{eq:cq_SGLPE_rot}). We increase $\Omega$ slowly and use the final steady-state value for $\psi$, for a given value of $\Omega$, as the initial condition for the next value of $\Omega$.

\begin{figure}
    \centering
    \includegraphics[scale=0.14]{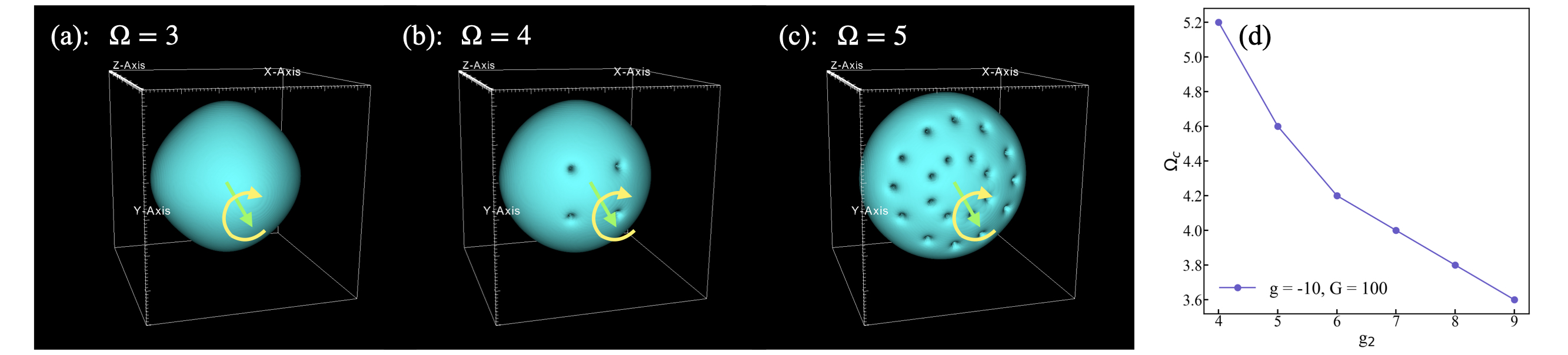}
    \caption{Contour plots of the density $|\psi|^2$, for a single rotating compact axionic object, which we obtain by solving the cq-SGLPE for (a)  $\Omega=3$, (b)  $\Omega=4$, and (c)  $\Omega=5$. Vortices appear once $\Omega=\Omega_c$, a critical angular speed. The $z$ axis of rotation is indicated by the green arrow. (d) Plot of $\Omega_c$ versus $g_2$ [adapted from Ref.~\cite{Shukla_PhysRevD.109.063009}, with permission from the APS].}
    \label{fig:axion_rotation}
\end{figure}

Figures~\ref{fig:axion_rotation} (a)-(c) show plots of $|\psi({\bf x},t)|^2$, for the rotating axion star [the green arrow indicates the rotation axis $(z)$], from our DNS of Eq.~\eqref{eq:cq_SGLPE_rot} for $g = -15$, $g_2=8$, $G = 100$, and (a) $\Omega = 3$, (b) $\Omega = 4$, and (c) $\Omega = 5$. Vortices nucleate once $\Omega > \Omega_c \gtrsim 3$, and their number density increases with $\Omega$ [Figs.~\ref{fig:axion_rotation} (b) and (c)]; and $\Omega_c$ decreases as $g_2$ increases [Fig.~\ref{fig:axion_rotation} (d)]~\footnote{See Ref.~\cite{PRA_omega_c_vs_g}, for similar results for a trapped BEC, without the $G$ and $g_2$ terms, where it is found that the critical angular speed decreases as the repulsive interaction $g$ between the bosons increases.}. If with $g <0$ and $G$, $\Omega_c$ becomes so  high vortices do not form  [Fig.~\ref{fig:axion_rotation} (d)].

\begin{figure}
    \centering
    \includegraphics[scale=0.09]{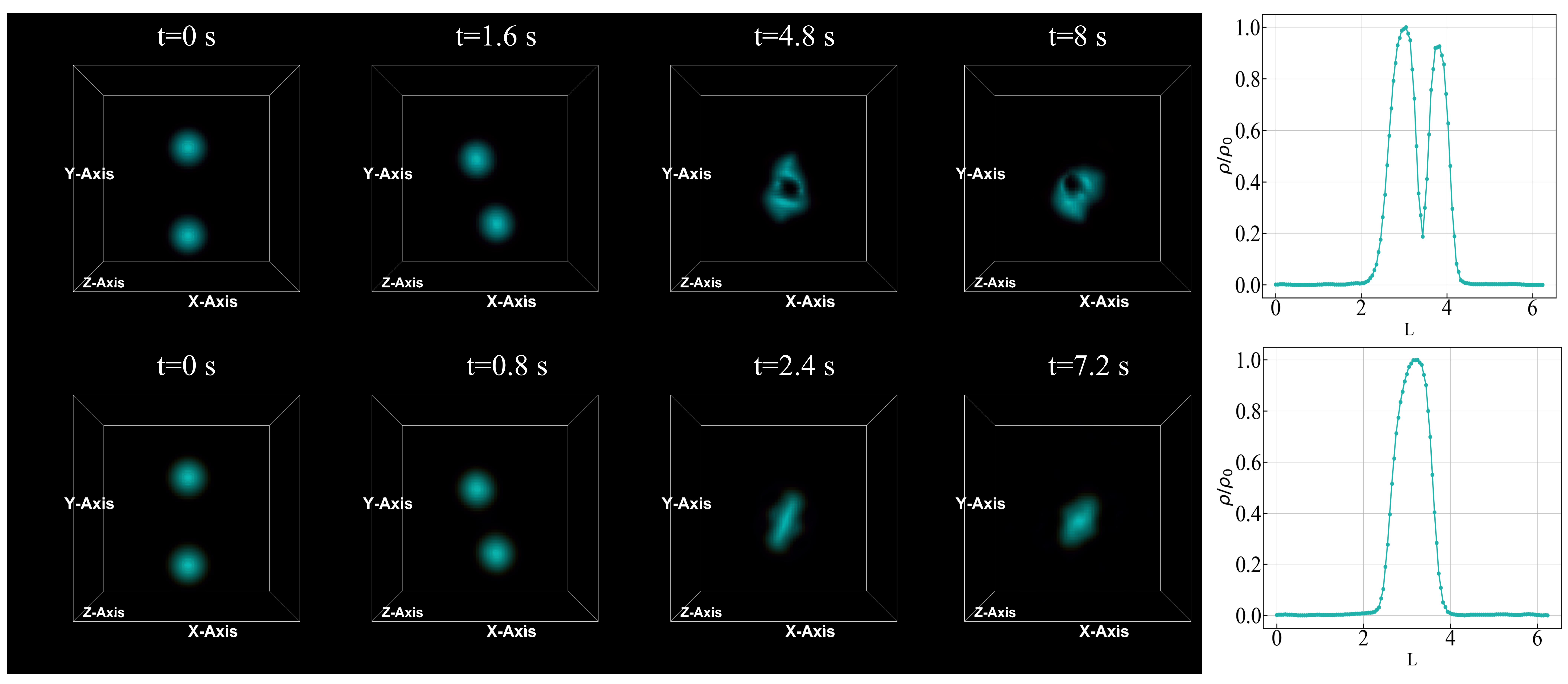}
    \caption{Volume plots of density $|\psi|^2$ for a rotating binary axionic system obtained by using the cq-GPPE, and for and relative phase $\Delta =\pi$ (first row) and $\Delta =0$ (second row). The last column shows plots of the density variation along the line passing through the center of the last configuration of the collapsed, rotating axionic system [adapted from Ref.~\cite{Shukla_PhysRevD.109.063009}, with permission from the APS].}
    \label{fig:gpe_binary_star2}
\end{figure}

To consider a rotating axionic binary system, we begin with two spherical collapsed axionic objects,with the same mass, $N_1=N_2=N/2$, where $N$ is the number of axions, separated by a distance $d$, along the $y$-direction; furthermore, the initial velocities ${\bf v_1} = (v,0,0)$, ${\bf v_2} = (-v,0,0)$, in the $x$-direction and
\begin{eqnarray}
\psi_b({\bf x},t=0) &=& f(|{\bf x - x_0}|)e^{i\bf v_1\cdot x}+ f(|{\bf x + x_0}|)e^{i{\bf (v_2\cdot x} + \Delta \phi)}\,, 
\label{eq:binary_ini}
\end{eqnarray}
where ${\bf x_0} = (0,d/2,0)$ and $\Delta \phi$ is a phase. We obtain the function $f(|\bf x|)$ using the initial condition {\bf IC2} in Ref.~\cite{Shukla_PhysRevD.109.063009}, which converges rapidly to the stationary state of the cq-GPPE~(\ref{eq:cq_TGPPE}).
%We choose two axionic compact objects that have the same mass, $N_1=N_2=N/2$, where $N$ is the number of axions. 
We then evolve Eq.~(\ref{eq:cq_TGPPE}) in time, for $v = 0.5$, and $ g=-0.5,\, g_2=0.005,\,$  and $G=2.0$.
Our results are summarised in Fig.~\ref{fig:gpe_binary_star2} [see Ref.~\cite{Shukla_axion_PhysRevD.109.063009} for details]. The plots of $|\psi({\bf x},t)|^2$ are for $\Delta \phi = \pi$ (dark top panel) and $\Delta \phi = 0$ (dark bottom panel); the merger occurs more slowly in the first case than in the second; after the merger, the single collapsed object rotates with a finite angular momentum and quantized [if this angular momentum is sufficiently high and $g_2$ is large as in the top panel of Fig.~\ref{fig:gpe_binary_star2}]. If we plot the density along one direction, such a
vortex leads to a large dip in the density  [last column of Fig.~\ref{fig:gpe_binary_star2}].
% can also be visualized by plotting the density variation along one direction for the last configuration of the collapsed, rotating
% axionic object. We show such plots in the last column of Fig.~\ref{fig:gpe_binary_star2}. The large dip in the density, near the middle, indicates a vortex when $\Delta \phi=\pi$; we do not observe such a dip if $\Delta \phi=0$.

\section{Superfluids inside neutron stars}
\label{sec:NS}

% In the laboratory, the transition temperature for superfluidity is very low (e.g. $T_{\lambda}=2.17\ {\rm K}$ for liquid $^4 {\rm He}$). However, there are regions in the universe where the temperature can be very low, such as the interior of a neutron star, but the temperature is typically of the order of tens of degrees Kelvin. 
It was first pointed out by Migdal~\cite{MIGDAL1959655} and Ginzburg~\cite{Vitalii_L_Ginzburg_1969} that superfluidity can occur in neutron stars whose
interiors have extremely large densities. 
%To determine the superfluidity conditions for such systems, we calculate 
Recall that the transition temperature for a BEC is 
\begin{eqnarray}
    T_{\rm C} = \frac{3.31}{g^{3/2}}\frac{\hbar^2}{mk_{\rm B}}n^{2/3}\,,
\end{eqnarray}
where $g=2S+1$ is the spin degeneracy, $m$ is the mass of a boson, and $n$ is the number density. For a plasma of Helium nuclei, in stellar interiors, superfluidity occurs 
for $T< T_{\rm C}$, with the transition temperature
\begin{eqnarray}
    T_{\rm C} = 3.31\frac{\hbar^2}{M^{5/3}_{\rm He}k_{\rm B}}\rho^{2/3}\approx 11 \rho^{2/3} \,,
\end{eqnarray}
where $M_{\rm He}$ is the mass of the  helium nuclei, and $\rho = M_{\rm He}n$ is the density of the gas. 
%Superfluidity of these helium nuclei will occur at $T< T_{\rm C}$. 
However, at this density $\rho$ and $T< T_{\rm C}$, Helium atoms must be fully ionized and, at the same time, they should not be crystallized~\cite{Vitalii_L_Ginzburg_1969}. %~\footnote{The number of protons is sufficient to maintain charge neutrality.}.
For sufficiently strong compression, materials crystallize; the melting temperature, for a given density in a plasma of Helium nuclei, can be estimated to be~\cite{Vitalii_L_Ginzburg_1969} $ T_{\rm m} \approx 3 \times 10^3 \rho^{1/3}$. 
%\begin{eqnarray}
%    T_{\rm m} \approx 3 \times 10^3 \rho^{1/3}\,.
%\end{eqnarray}
We see that $T_{\rm C}< T_{\rm m}$ for densities $\rho \le 3 \times 10^7\ {\rm g \ cm^{-3}}$, so superfluidity can occur at densities $\rho \ge 3 \times 10^7\ {\rm g \ cm^{-3}}$, with $T_{\rm C} \ge 10^6\ {\rm K}$~\cite{Vitalii_L_Ginzburg_1969}. 

Such extreme densities and transition temperatures can occur in the interior of a neutron star, where neutrons constitute the major part ($\simeq 95 \% $ of the total mass). These neutrons form Cooper pairs and undergo Bose-Einstein condensation. There are also a few protons ($\simeq 5 \%$ of the total mass) inside a neutron star. Shortly after the suggestion of neutron superfluidity, Baym, Pethik, and Pines~\cite{Baym_1969Natur} pointed out that protons could also form Cooper pairs, leading to Type II superconductivity inside a neutron star. These proton Cooper pairs generate their own magnetic field; the magnetized rotating neutron star is a pulsar.
%\begin{figure}[!hbt]
%    \centering
%   \includegraphics[scale=0.2]{figures/Pulsar_emission.jpeg}
%    \caption{This schematic figure illustrates a magnetized rotating pulsar with radiation beams emitted from its poles: {\bf (a)} when the radiation beam is oriented away from Earth, resulting in no detectable signal, and {\bf (b)} when the radiation beam is directed towards Earth, producing a detectable pulse.}
%    \label{fig:pulsar_emission}
%\end{figure}

\begin{figure}[!hbt]
    \centering
    \includegraphics[scale=0.1]{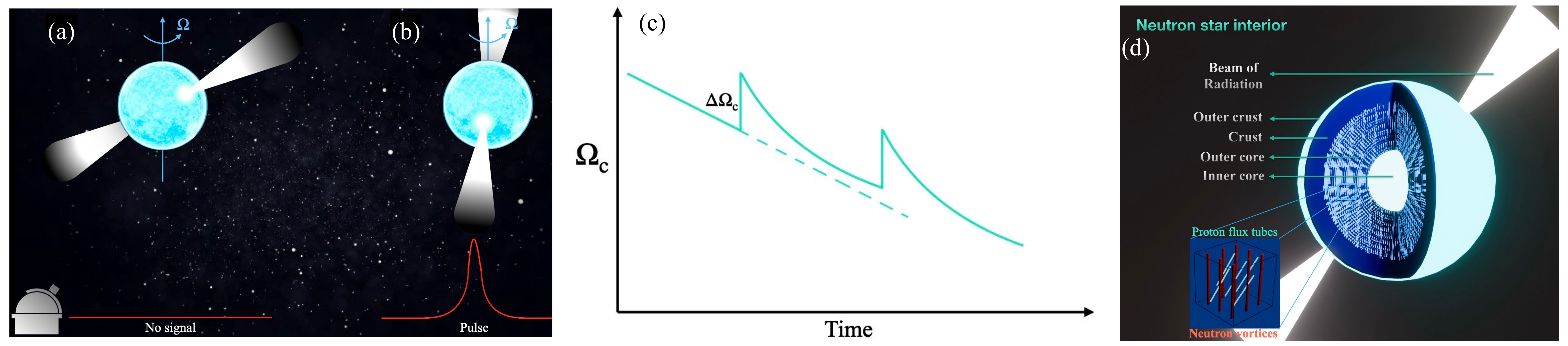}
    \caption{(a)-(b): This schematic figure illustrates a magnetized rotating pulsar with radiation beams emitted from its poles: {\bf (a)} when the radiation beam is oriented away from Earth, resulting in no detectable signal, and {\bf (b)} when the radiation beam is directed towards Earth, producing a detectable pulse. (c): The figure presents the rotational frequency ${\bf \Omega}_{\rm c}$ of a pulsar's crust over time $t$, highlighting $\Delta {\bf \Omega}_{\rm c}$, the abrupt change in frequency commonly referred to as glitches. (d): A schematic diagram of the interior of a pulsar. Light-blue \textit{inner core}, with ultra-dense matter where neutrons and protons break down into quarks and gluons; shaded blue-white \textit{outer core}, with a neutron superfluid and a proton superconductor [magnified view in the inset]; dark-blue crust that has a lattice structure with heavy atomic nuclei and free neutrons and electrons; 
    % The neutrons in the crust exist in the form of a superfluid that is threaded by vortices. The glowing white region, often called the outer crust, comprises atomic nuclei and free electrons. 
    white cones show radiation beams from the poles of the pulsar [adapted from Ref.~\cite{Shukla_PRD_2024}, with permission from the APS].}
    \label{fig:crust_frequency}
\end{figure}

The rotation and magnetic axes of a pulsar are not aligned, resulting in electromagnetic radiation being emitted at an angle to the rotation axis, as depicted schematically in Fig.~\ref{fig:crust_frequency} (a). This radiation is observed when the beam points toward Earth, producing a recorded pulse [Fig.~\ref{fig:crust_frequency} (b)], like the light emitted from a lighthouse sweeping a cone in space [hence the name \textit{lighthouse effect}]. These pulses occur at very regular rotational frequencies. Occasionally, there are sudden changes (increases or decreases) in the rotational frequencies of some pulsars; these are known as glitches.

Since the observation of pulsar glitches \cite{Radhakrishnan_1969Natur, Manchester2017PulsarG}, there have been several attempts to understand them. One of the most intriguing mechanisms behind these glitches is related to the existence of neutron superfluidity inside pulsars. As a pulsar rotates, the superfluid inside it also rotates by forming an array of quantized vortices~\cite{Muslimov_1985Ap&SS}, which carry all the angular momentum stored within. When a pulsar slows down as a result of electromagnetic radiation, the superfluid inside it does not experience this slowdown because of its zero viscosity. This creates a differential rotation between the interior superfluid and the outer crust of the pulsar. Because of this differential rotation, some of the quantized vortices move outward and transfer their angular momentum to the crust, causing a sudden change in the rotational frequency, i.e., a glitch.

Quantum vortices naturally arise in the Gross-Pitaevskii (GP) description of a superfluid. Therefore, we can model pulsar glitches by coupling the GP equation with an outer crust potential to which the quantized vortices can be pinned. Such modelling has been attempted using the 2D GP equation in a rotating container coupled with a pinning potential of the crust~\cite{Warszawski_2011, Warszawski_2012}. Recent work from our group has extended these efforts by using the 3D GPPE equation along with a crust potential for pinning~\cite{Verma_PhysRevResearch.4.013026}; self-gravity, which is crucial inside dense neutron stars, is included via the Poisson equation.

The neutron superfluid inside a pulsar can be described by a complex wavefunction $\psi_{\rm n}$, which satisfies the GP equation in the weak interaction limit. When combined with rotation, the GP equation is given by~\cite{Drummond_2017,Drummond_2017_b}
\begin{eqnarray}
    i\hbar\frac{\partial\psi_{\rm n}}{\partial t}=-\frac{\hbar^2}{2m_{n}}\nabla^2 \psi_{\rm n}+g|\psi_{\rm n}|^2 \psi_{\rm n}+i\hbar ({\bf \Omega}\times {\bf r})\cdot \nabla \psi_{\rm n}+V_{\rm ext} \psi_{\rm n}+V_{\rm pin}\psi_{\rm n}\,,
    \label{eq:GPPE_pulsar}
\end{eqnarray}
where $m_{\rm n}$ is the mass of a neutron Cooper pair, $g=\tfrac{4\pi a_s\hbar^2}{m_{\rm n}}$ with $a_s$ being the scattering length, and ${\bf \Omega}$ is the rotational velocity. The above Eq.~\eqref{eq:GPPE_pulsar} has been used to study pulsar glitches in an external rotating potential $V_{\rm ext}$, such as the harmonic trap. Here,  $V_{\rm pin}$ is the crust pinning potential to which vortices can be pinned. 
% One aspect missing from earlier models is the incorporation of self-gravity. A neutron star consists of a large number of particles that interact gravitationally, forming a gravitationally bound object. To account for the gravitational interaction between neutrons, we must couple the GP equation with a self-gravitating potential $\Phi$. The use of $\Phi$ results in a gravitationally bound object instead of one trapped harmonically.
Recent work by Verma, \textit{et al.}~\cite{Verma_PhysRevResearch.4.013026} uses the gravitational potential $V_{\rm ext } =m_{\rm n}\Phi$, which satisfies the Poisson equation
\begin{eqnarray}
    \nabla^2\Phi = 4\pi G(m_{\rm n}|\psi_{\rm n}|^2-\rho_{\rm bg})\,,
   \label{eq:poisson_equation}
\end{eqnarray}
where $G$ is Newton's gravitational constant; the subtraction of the mean background density $\rho_{\rm bg}$ is often called the Jeans swindle~\cite{Falco_2013}; it can be justified by defining a Newtonian cosmological constant~\cite{KIESSLING_2003}. 
% The GP equation coupled with the gravitational potential $\Phi$ is called the Gross-Pitaevskii-Poisson equation (GPP).
As the system evolves according to Eqs.~\eqref{eq:GPPE_pulsar} and \eqref{eq:poisson_equation}, it generates quantum vortices beyond a critical value ${\bf \Omega}_{\rm cr}$. The motion of these quantum vortices carries the angular momentum ${ L}_z$. When the pulsar slows down, these quantum vortices unpin and transfer their angular momentum to the crust. This transfer of angular momentum from the superfluid to the crust causes sudden changes in the rotational frequency of the crust, as shown schematically in Fig~\ref{fig:crust_frequency}(c) by $\Delta \Omega_{\rm c}$.
%\begin{figure}[!hbt]
%    \centering
%    \includegraphics[scale=0.25]%
%\caption{The figure presents the rotational frequency ${\bf \Omega}_{\rm c}$ of a pulsar's crust over time $t$, highlighting $\Delta {\bf \Omega}_{\rm c}$, the abrupt change in frequency commonly referred to as glitches.}
 %   \label{fig:crust_frequency}
%\end{figure}

We have only considered neutron superfluidity, so far, and neglected the proton superconductor component, which comprises $\simeq 5\%$ of the total mass in a neutron star and can form a Type II superconductor~\cite{Baym_1969Natur} with flux tubes. These flux tubes carry a strong magnetic field and affect the motion of neutron vortices, which in turn influences pulsar glitches. Some earlier studies, which have included the proton superconductor component coupled to the neutron superfluid~\cite{Drummond_2017,Drummond_2017_b,Thong_2023}, either employed a static Ansatz for the proton Cooper pairs in a harmonic trap~\cite{Drummond_2017,Drummond_2017_b} 
or used~\cite{Thong_2023} use the time-dependent Ginzburg-Landau equation (TDGLE)~\cite{Schmid_1966PKM,tinkham2004introduction} for the proton superconductor, \textit{but with a constant vector potential} ${\bf A}$. In certain high-frequency regimes, the full Maxwell equations for ${\bf A}$ are required. Recently, we have developed a fully coupled model that includes the neutron superfluid, proton superconductor, Maxwell's equations for ${\bf A}$, and the Poisson equation for self-gravity~\cite{Shukla_PRD_2024}.

In a pulsar, neutrons form Cooper pairs that lead to a superfluid~\cite{MIGDAL1959655,Rezzolla:2018jee}. At temperatures below the transition temperature $T_{\lambda}$, these Cooper pairs lead to a Bose-Einstein condensate (BEC), characterized by a macroscopic complex wavefunction $\psi_n$. Proton Cooper pairs, which also form in a pulsar, yield a Type II superconductor with an Abrikosov flux lattice~\cite{Baym_1969Natur}. This superconducting system can be described by the complex wavefunction $\psi_p$, coupled to a vector potential $\bf A$ and self-gravitating potential $\Phi$. We use the GPPE for neutron Cooper pairs and the real-time Ginzburg-Landau-Poisson equation (RTGLPE) for proton Cooper pairs. We refer the reader to Eqs.(19) and (20) in our earlier work~\cite{Shukla_PRD_2024} for the GPPE and RTGLPE [not given here because of length restrictions],
which are coupled with the vector potential ${\bf A}$, and the Poisson equations for the gravitational potential $\Phi$, and the electric scalar potential $\phi$ [Eqs.(21), (22), and (23) in Ref.~\cite{Shukla_PRD_2024}].
In addition, we must account for direct density-density and current-current interactions between the neutron and proton subsystems. The density-density interaction $g_{np}$ causes the cores of neutron vortices and proton flux tubes to attract ($g_{np}<0$) or repel ($g_{np}>0$), thus affecting the movement of vortices inside the condensate. The current-current interaction, causes an entrained proton current around neutrons by producing an induced magnetic field around neutron vortices. Both of these interactions play important roles in the spatiotemporal evolution of neutron vortices and, thereby, the statistics of glitches in our model pulsars.
%\begin{figure}[!htb]
%    \centering 
%    \includegraphics[scale=0.18]{figures/neutron_star_lattice_4.jpeg}
%    \caption{A schematic diagram of the interior of a pulsar (magnetized neutron star). The light-blue luminous central region represents the \textit{inner core}, characterized by ultra-dense matter where neutrons and protons break down into quarks and gluons. Surrounding this core lies the \textit{outer core} (shaded blue-white), composed of a neutron superfluid and proton superconductor, with neutron-superfluid vortices and proton-superconductor flux tubes, respectively [magnified view in the bottom-left inset]. The dark-blue crust has a crystalline lattice structure (not shown) and consists of heavy atomic nuclei and free neutrons and free electrons. The neutrons in the crust exist in the form of a superfluid that is threaded by vortices. The glowing white region, often called the outer crust, comprises atomic nuclei and free electrons. The white conical regions show radiation beams emerging from the poles of the pulsar. }
%    \label{fig:neutron_star_cartoon}
%\end{figure}

% The crust, just above the outer core of a neutron star [Fig.~\ref{fig:crust_frequency}(d)], contains heavy nuclei arranged in a crystalline lattice. 
Neutron vortices become pinned to the lattice sites in the crust [Fig.~\ref{fig:crust_frequency}(d)] and co-rotate with it. 
As it spins down, the superfluid within remains unaffected, the differential rotation causes vortices to unpin from their pinning sites, momentum is transferred to the crust and results in glitches. 
%Flux tubes are also anchored to the crust by the strong magnetic field; and there is a depletion of the proton Cooper pairs. 
% In our model~\cite{Shukla_PRD_2024}, the collapsed neutron and proton condensates contain vortices and flux tubes that are located away from the boundary of our cubical simulation domain. The crust potential lies just above the condensate, with the magnetic field inside the flux tubes passing through the crust anchoring them. 

We can model~\cite{Shukla_PRD_2024} the crust using a Gaussian potential $V_{\theta}$ modulated by equally spaced pinning centres. In the absence of the crust potential $V_{\theta}$, Eqs.~(19)-(20) in Ref.~\cite{Shukla_PRD_2024} govern the interplay between neutron-superfluid vortices and proton-superconductor flux tubes [see the outer-core region in Fig.~\ref{fig:crust_frequency}(d)]. At the level of our minimal model for pulsars, the dynamics of this crust is characterised by a single polar angle $\theta$~\cite{AK_verma_2022,Shukla_PRD_2024} that evolves as follows:
\begin{eqnarray}I_c\ddot{\theta}={\rm F}_s -\delta \dot{\theta}\,,
    \label{eq:crust_pot_eq}
\end{eqnarray}
%\begin{equation}
%\begin{aligned}
%I_c\frac{d^2\theta}{dt^2} &= \frac{1}{N_n}\bigg(\int d^3x\partial_{\theta}V_{\theta}|\psi_n|^2 +\frac{n_n}{n_p}\int d^3x\partial_{\theta}V_{\theta}|\psi_p|^2\bigg)-\delta \frac{d\theta}{dt}\,;\\
%V_{\theta} ({\bf r}_p) &= V_0\exp \left[-\frac{(|{\bf r}_p|-r_{\rm crust})^2}{(\Delta r_{\rm crust})^2}\right] \tilde{V}(x_{\theta},y _{\theta})\,; 
% \end{aligned}
% \label{eq:crust_pot_eq}
%\end{equation}
where $I_c$ is the moment of inertia of the crust, $\theta$ represents the angular rotation of the crust about the rotation axis, ${\rm F}_s$ is the force of the superfluid on the crust, incorporating the crust potential $V_\theta$, and the deceleration of the crust is controlled by the friction coefficient $\delta$.
%$N_n = \int |\psi_n|^2 d^3 x$ is the total number of neutron Cooper pairs, $n_n/n_p$ is the ratio of the number densities of neutrons and protons, and the slowing down of the crust is controlled by the friction coefficient $\delta$. The first two terms on the right-hand side of upper Eq.~\eqref{eq:crust_pot_eq} couple the crust to the superfluid and superconductor, respectively. These terms ensure that the superfluid and superconductor act on the crust. The last term on the right-hand side of upper Eq.~\eqref{eq:crust_pot_eq} represents the friction, which slows down the crust and creates a differential rotation between it and the superfluid. The evolution Eq.~\eqref{eq:crust_pot_eq} for the crust potential can be written in the compact form $I_c\ddot{\theta}={\rm F}_s -\delta \dot{\theta}$, where ${\rm F}_s$, the force of the superfluid on the crust, is given by the term in parentheses in the first line of Eq.~\eqref{eq:crust_pot_eq}. In our minimal model the crust potential $V_{\theta}$ is a function of a single polar angle $\theta$ whose dynamics is given by Eq.~(\ref{eq:crust_pot_eq}). 

To study the interactions of the crust, neutron-superfluid, and proton-superconductors, we use Eqs.~(19)-(20) in Ref.~\cite{Shukla_PRD_2024} along with Eq.~(\ref{eq:crust_pot_eq}); we use initial conditions from equilibrium states that we obtain from the imaginary-time GPPE and the RTGLPE [see Ref.~\cite{Shukla_PRD_2024} for details]. 
% For the purpose of illustration, we consider $\Theta=0$, i.e., the angle between the rotation axis and the magnetic field is zero; and the neutrons and protons interact only through the gravitational potential via the Poisson equation. 

\begin{figure}[!htb]
    \centering
    \includegraphics[scale=0.2]{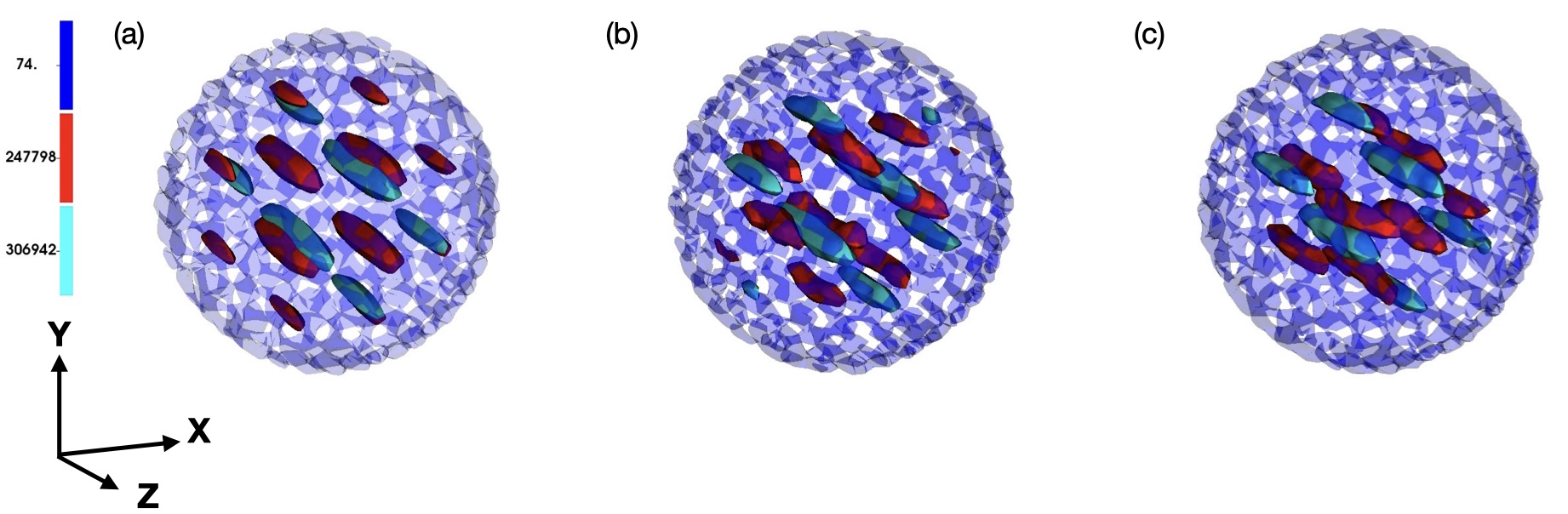}
    
    \caption{One-level contour plots of the crust potential together with the neutron vortices (in red) and proton flux tubes (in cyan) at three different times in {\bf (a)}, {\bf (b)}, and {\bf (c)} obtained by using the real-time GPPE and  RTGLE [Eqs.(19)-(20) in Ref.~\cite{Shukla_PRD_2024}]. Both neutron and proton subsystems rotate with an angular velocity ${\bf \Omega} = \Omega \hat{z}$, where $\Omega=4.0$; and ${B}_{\rm ext}=4.0$, which is along the $z$-axis [adapted from Ref.~\cite{Shukla_PRD_2024}, with permission from the APS].}
    \label{fig:gpe_3D_npv_crust}
\end{figure}

In Figs.~\ref{fig:gpe_3D_npv_crust}(a)-(c), we show the crust potential (blue), neutron-superfluid vortices (red), and proton-superconductor flux tubes (cyan), at three times [and for $\Theta=0$]. Initially we have $12$ neutron vortices and $6$ proton flux tubes [Fig.~\ref{fig:gpe_3D_npv_crust}(a)]. Given our parameters, 
% the proton-superconductor flux tubes are more effectively pinned by the crust potential than neutron-superfluid vortices, partly because of the anchoring of the flux tubes to the strong external magnetic field ${\bf B}_{\rm ext}$, so, 
the first number reduces to 6, whereas the number of flux tubes remains unchanged [Figs.~\ref{fig:gpe_3D_npv_crust}(a)-(c) and Ref.~\cite{Shukla_PRD_2024}]. 
% Both superfluid vortices and proton-superconductor flux tubes undergo differential rotation because of the friction coefficient $\alpha$ in Eq.~(\ref{eq:crust_pot_eq})). 

The time dependence of the crust angular momentum ${\rm J}_c \equiv I_c d\theta/dt$, where $I_c$ is the moment of inertia, is complicated because of the competition between the angular momentum in the neutron-superfluid vortices and the friction. The ejection of such a vortex results in the transfer of its angular momentum to the crust. Furthermore, ${\rm J}_c$ is affected by the Poisson-equation-induced gravitational attraction between the neutron and proton subsystems, by virtue of which some neutron vortices remain close to proton-superconductor flux tubes. Eventually, there as an effective stick-slip evolution for ${\rm J}_c$ so we see glitches, whose statistical properties are like those observed in various pulsars~\cite{Radhakrishnan_1969,reichley1969observed,manchester_2017,Verma_PhysRevResearch.4.013026}.

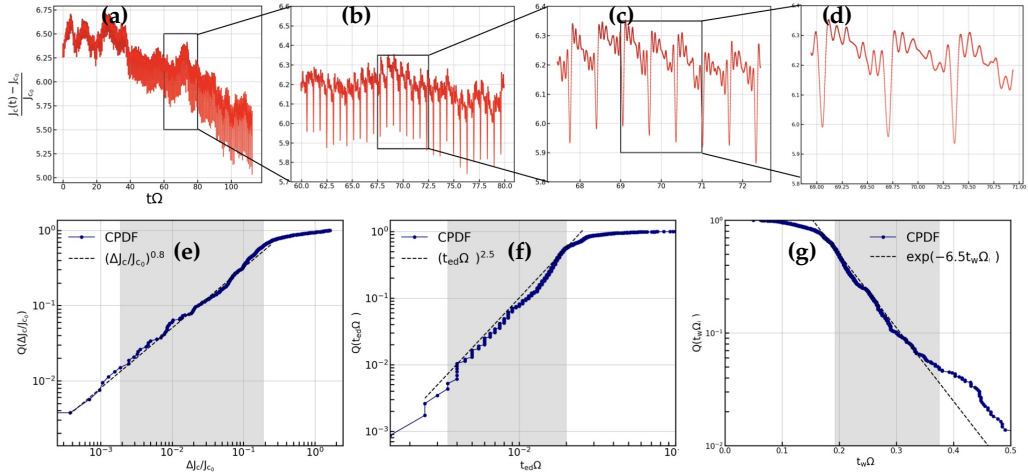
\begin{figure}[!htb]
    \centering
    \includegraphics[scale=0.2]{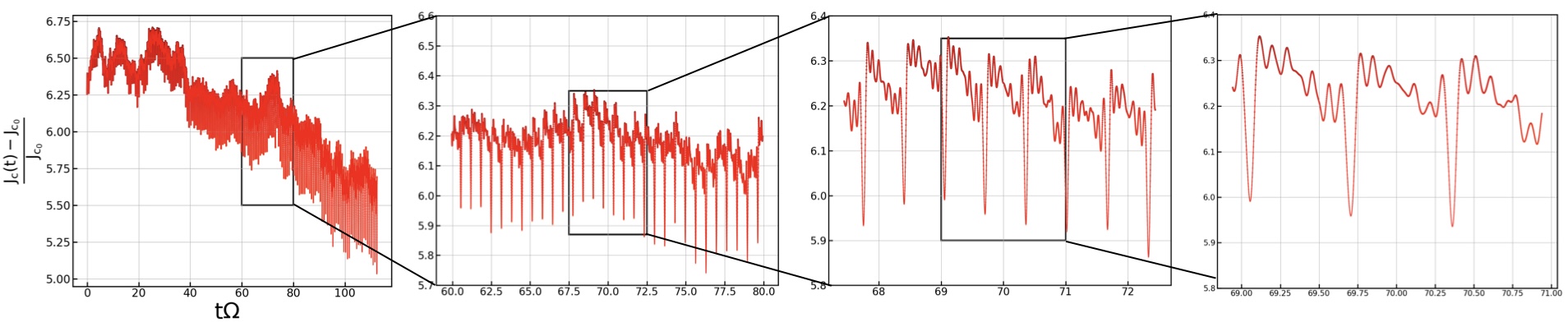}
    \put(-350,70){ \bf (a)}
    \put(-260,70){ \bf (b)}
    \put(-160,70){ \bf (c)}
    \put(-80,70){ \bf (d)}

    \includegraphics[scale=0.2]{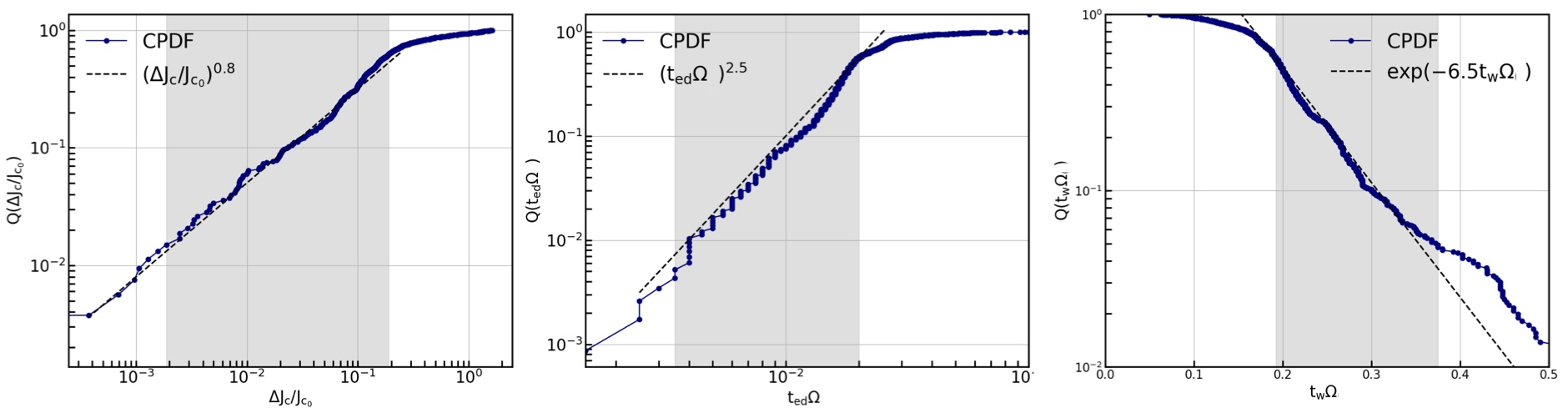}
    \put(-320,80){ \bf (e)}
    \put(-195,80){ \bf (f)}
    \put(-90,80){ \bf (g)}
    \caption{(a) Plot of the crust angular momentum $\rm (J_c-J_{c_0})/J_{c_0}$ versus time. (b), (c), and (d) display enlarged versions of the boxed regions in the preceding plots. Log-Log plots of CPDFs: (e) $Q(\Delta J_c/J_{c_0})$ and (f) $Q(\rm t_{ ed}\Omega)$; and (g) semilog plot of the CPDF $\rm Q(\rm t_{ w}\Omega)$ (see text). $\rm J_{c_0}$ and $\rm \Omega$ are the initial angular momentum and initial angular velocity of the crust, respectively [adapted from Ref.~\cite{Shukla_PRD_2024}, with permission from the APS].}
    \label{fig:gpe_3D_crust_angmom}
\end{figure}

We now examine the analogues of pulsar glitches in our model~\cite{Shukla_PRD_2024}, by following Ref.~\cite{AK_verma_2022}. We
begin with the time series of $ ({\rm J}_c-{\rm J}_{c_0})/{\rm J}_{c_0}$  
[Fig.\ref{fig:gpe_3D_crust_angmom} (a)], which exhibits characteristic features of Self-Organized Criticality (SOC)~\cite{Bak_1987,Jensen_1998,Donald_Turcotte_1999,Melatos_2008}. We have explored these features earlier in the context of gravitationally collapsed boson stars~\cite{AK_verma_2022}. Figures~\ref{fig:gpe_3D_crust_angmom}(b)-(d) present expanded views of specific segments [indicated by black boxes] of the time series in Fig.\ref{fig:gpe_3D_crust_angmom}(a); we see that the crust can either lose angular momentum to the superfluid or gain angular momentum from it, because of the stick-slip dynamics mentioned above. 

Moreover, we calculate cumulative probability distribution functions (CPDFs that we denote generically by $Q$ below) of $\Delta  {\rm J}_c$, $t_{ed}$,
and $t_{w}$, which are, respectively, the event size $\Delta  {\rm J}_c$ [the difference between successive minima and maxima in ${\rm J}_c$], the event-duration time $t_{ed}$ [the time difference between successive minima and maxima of $ {\rm J}_c(t)$], and the waiting time $t_{w}$ [the time between successive maxima in $ {\rm J}_c(t)$].  The first two 
of these CPDFs show power-law tails, whereas the last displays an exponential tail [cf. Ref.~\cite{AK_verma_2022}]:
Specifically, $Q(\Delta {\rm J}_c/{\rm J}_{c_0})\sim (\Delta  {\rm J}_c/{\rm J}_{c_0})^{\beta}$, in the gray region of Fig.~\ref{fig:gpe_3D_crust_angmom}(e), so the associated probability distribution function (PDF) $P(\Delta {\rm J}_c/{\rm J}_{c_0})\sim (\Delta {\rm J}_c/{\rm J}_{c_0})^{\beta-1}$; for our run, the exponent $\beta = 0.86\pm 0.15$. Similarly, $Q(t_{ed}\Omega)\sim (t_{ed}\Omega)^{\gamma_t}$ in the gray region of Fig.~\ref{fig:gpe_3D_crust_angmom}(f), with 
$\gamma_t=2.5\pm 0.2$ for our run. The CPDF $Q(t_{w}\Omega)\sim \exp(-6.5 t_{w}\Omega)$ [Fig.~\ref{fig:gpe_3D_crust_angmom}(f)]. The forms of these CPDFs are similar to those observed for various pulsars [cf. Ref.~\cite{AK_verma_2022}, which uses the GPPE]; and the for $\beta$ and $\gamma$ are close to those in some observations [e.g., PSR J 1825-0935 has glitch-size-PDF exponent $\simeq 0.36 $ ]~\cite{Melatos_2008}. Reference~\cite{AK_verma_2022} has found that $\beta$ depends on $\Omega$; we find that it also depends on $B_{\rm ext}$.

Our minimal model for a pulsar has been extended recently to study magnetars~\cite{shukla2025dynamics}.

\section{Conclusion}

We have presented an overview of the Gross-Pitaevskii-Poisson equation (GPPE) framework for self-gravitating 
superfluid systems, including boson and axion stars and dark-matter haloes. Our overview, based principally on the work from our group~\cite{Verma_PhysRevResearch.4.013026,AK_verma_2022,AK_Verma_PhysRevResearch.3.L022016,Shukla_axion_PhysRevD.109.063009,Shukla_PRD_2024,shukla2025dynamics,shukla_2025_PRR}, describes how this framework can be used to develop minimal models for neutron stars and for pulsars and their glitches. We examine vortices in the neutron superfluid inside these stars and the flux tubes in the proton-superconductor subsystem. Our coupled model includes the neutron superfluid, proton superconductor, the Maxwell equations for the vector potential ${\bf A}$, and the Poisson equation for self-gravity; the integration of these elements has not been attempted so far. Nevertheless, this is a minimal model, for it does not include all the complexities of neutron stars and dark-matter haloes. We give a brief discussion below of the limitations of our model. In particular, we note that the GPPE framework must be applied with care when we consider systems as different as the interiors of neutron stars and haloes around galaxies. 

Dark-matter haloes (DMHs) around galaxies are considered to be made up of bosons or axions, which are spin-0 particles. The masses of these bosons or axions should be $m\simeq 10^{-22}eV/c^2$ for them to condense on galactic scales [to form a DMH of size $\simeq 1\ kpc$], so the classical treatment of DMHs, within the GPPE framework, seems to be applicable at the galactic scale. If the DHM around galaxies has a radius $R$ and total mass $M$, the virial theorem gives the velocity $v^2 = \tfrac{GM}{R}$. The  gravitational collapse of the bosons in the DMH can be prevented because of the Heisenberg uncertainty principle $\Delta p \Delta r\sim \hbar$, which introduces the smallest length sacle $r_0$ in the system. Using $R\gg r_o$, we get $v\ll GMm $~\cite{Sin_PhysRevD_1994}. A nonrelativistic treatment is justified when $GMm \ll 1$; for bosons with a mass $m\simeq 10^{-22}eV/c^2$ the relevant mass scale turns out to be $M\simeq 10^{10}M_{\odot}$, where $M_{\odot}$ is a solar mass. Furthermore, the self-interaction between bosons or axions affects the density distribution of the DMH around galaxies.  The self-interaction between these objects comes from the scattering length $a_s$, which is positive for bosons ($a_s>0$) and negative for axions ($a_s<0$). 

Both the cases of repulsive ($a_s>0$)~\cite{Chavanis_PhysRevD.84.043531,AK_Verma_PhysRevResearch.3.L022016} and attractive ($a_s<0$)~\cite{Chavanis_att_PhysRevD.94.083007,Shukla_axion_PhysRevD.109.063009} self-interactions can be treated using the GPPE. For the repulsive case, the GPPE has the usual cubic nonlinear term, and the bosons form a stable condensate by balancing the repulsive self-interaction force and repulsive quantum pressure force against the attractive gravitational force. As we have discussed, for an attractive self-interaction ($a_s<0$), to avoid collapse, we  must include the positive quintic  nonlinear term in the GPPE model for axion stars~\cite{Shukla_axion_PhysRevD.109.063009,Chavanis_axion_PhysRevD.98.023009}.   

The application of the GPPE inside neutron stars: The BECs form because of neutron Cooper pairs [neutrons with opposite spins attract each other in a degenerate sea of neutrons]. Such states inside neutron stars occur at very high densities $\rho\simeq 10^{14} g\cdot cm^{-3}$ and so have very high transition temperatures $T_C\simeq 10^{11}\ K$~\cite{MIGDAL1959655}. 
%So, a classical treatment of the GPP using the Poisson equation is not applicable to the given length scales of neutron stars. 
However, a lot of physics can be extracted using the GPPE formulation inside neutron stars. As we have shown, one such example is the explanation of glitches found in pulsars~\cite{Radhakrishnan_1969,Boynton_1969}, which are magnetised neutron stars. Apart from neutrons, pulsars also consist of protons. These protons form Cooper pairs and exist in the form of superconductors~\cite{Baym_1969Natur}. Although they comprise $5\%$ of the total mass in a neutron star, protons affect the dynamics of the magnetic field inside pulsars and their glitches. We have demonstrated that the GPPE framework can be extended to include the dynamics of proton Cooper pairs by coupling it with the vector potential. 
% The interactions between the neutron and proton subsystems are essential to fully understand the behaviour of magnetic fields and glitches in pulsars, which can be explored by coupling the dynamics of the GPP and RTGL equations. 

\ack{S. Shukla and R. Pandit thank the Anusandhan National Research Foundation (ANRF), the Science and Engineering Research Board (SERB), and the National Supercomputing Mission (NSM), India, for support,  and the Supercomputer Education and Research Centre (IISc), for computational resources.}

%%%%%%%%%% Insert bibliography here %%%%%%%%%%%%%%

%\begin{thebibliography}{9}

%\bibitem{1} Allwood JM, Cullen JM. 2011 \textit{Sustainable materials:  with both eyes open}.
%Cambridge, UK: UIT Cambridge. See \href{http://www.withbotheyesopen.com}{http://www.withbotheyesopen.com}.

%\bibitem{2}  MacKay DJC. 2008  \textit{Sustainable energy:  without the hot air}.
% Cambridge, UK: UIT Cambridge. See \href{http://www.withouthotair.com}{http://www.withouthotair.com}.

%\bibitem{3} Gallman PG. 2011  \textit{Green alternatives and national energy strategy: the facts
% behind the headlines}.  Baltimore,\ MD: Johns Hopkins University Press.

%\bibitem{4} MacKay DJC. 2013.  Solar energy in the context of energy use, energy transportation, and
% energy storage. \textit{Proc. R. Soc. A} \textbf{371}.

%\end{thebibliography}

\bibliographystyle{ieeetr}
\bibliography{sample.bib}

\end{document}